\documentclass{elsarticle}
\usepackage[caption = false]{subfig}
\usepackage[multiple]{footmisc}
\usepackage{eurosym}

\usepackage{booktabs}

\usepackage[draft]{changes}

\usepackage{amsmath,amssymb,amsthm,mathrsfs,amsfonts,dsfont}

\usepackage{array}
\usepackage{epstopdf}
\usepackage{setspace}
\usepackage{algorithm,algorithmic}
\usepackage{natbib}
\usepackage{color}
\usepackage{rotating}
\usepackage{tikz}
\usepackage{subfig}
\usepackage[toc,page]{appendix}
\usepackage[font=small,skip=0pt]{caption}
\usepackage{eurosym}
\usepackage{hyperref}

\newcommand*\colvec[3][]{
	\begin{pmatrix}\ifx\relax#1\relax\else#1\\\fi#2\\#3\end{pmatrix}
}
\newcommand*\colvecfive[5][]{
	\begin{pmatrix}\ifx\relax#1\relax\else#1\\\fi#2\\#3 \\#4\\#5\end{pmatrix}
}

\theoremstyle{remark}

\theoremstyle{definition}

\numberwithin{table}{section}
\numberwithin{figure}{section}

\graphicspath{{figures/}}  

\journal{}
\begin{document}
	
	\begin{frontmatter}
	\title{Local and Global Balance in Financial Correlation Networks:\\ an Application to Investment Decisions}
		
		\author[Unimi]{Paolo Bartesaghi}
		\author[Unimib]{Rosanna Grassi\corref{cor}}
		\author[Unimib]{Pierpaolo Uberti}
			
		\cortext[cor]{\emph{Corresponding author. email: rosanna.grassi@unimib.it}}
		\address[Unimi]{University of Milano, Via Conservatorio 7, 20122 Milano, Italy}
		\address[Unimib]{University of Milano - Bicocca, Via Bicocca degli Arcimboldi 8, 20126 Milano, Italy}

		\begin{abstract}
The global balance is a well-known indicator of the behavior of a signed network. Recent literature has introduced the concept of local balance as a measure of the contribution of a single node to the overall balance of the network. In the present research, we investigate the potential of using deviations of local balance from global balance as a criterion for selecting outperforming assets. The underlying idea is that, during financial crises, most assets in the investment universe behave similarly: losses are severe and widespread, and the global balance of the correlation-based signed network reaches its maximum value. Under such circumstances, standard diversification (mainly related to portfolio size) is unable to reduce risk or limit losses. Therefore, it may be useful to concentrate portfolio exposures on the few assets—if such assets exist—that behave differently from the rest of the market. We argue that these assets are those for which the local balance strongly departs from the global balance of the underlying signed network. The paper supports this hypothesis through an application using real financial data. The results, in both descriptive and predictive contexts, confirm the proposed intuition.  
		\end{abstract}
		
	\end{frontmatter}

\section{Introduction}

Global balance of a signed network was introduced in \cite{Harary1953, Cartwright1956} to assess weather a network is structurally balanced. More recently, the concept of local balance has been proposed to measure the contribution of a single node to the overall balance of the network (for further details, see \cite{diazdiaz2024}).

A recent contribution proposes the global balance as a valuable measure of systemic risk in financial correlation networks \cite{bartesaghi2025a}.
In general, network theory is one possible approach to addressing systemic risk. The failure of one node can propagate and trigger defaults or distress across many others, depending on the network topology. The study by \cite{AllenGale2000} is a prominent contribution to the field. Since then, the literature on this topic has grown, exploiting different network tools. For example, various structural network indicators have been used as measures to assess risk contagion in the interbank system \cite{Battiston2012, Tabak2014, Bardoscia2017, Bongini2018}. Multi-layer financial networks have been proposed to model contagion propagation in bank-firm connections \cite{poledna2021, wang2025}. Dynamic network indicators are useful for tracing the dynamics of contagion across countries and sectors of the financial system \cite{FRANCH2024}.
In financial applications of signed networks, \cite{adhikari2025} proposes a discrete optimization model that reduces the asset selection problem to the desired size. Signed graphs have also been used to analyze correlated and anticorrelated asset returns in portfolio risk management and the structural balance of stock markets in \cite{Harary2002,ferreira2021}. The related literature is wide, and we refer the reader to \cite{pacelli2025} for a detailed treatment.

During financial crises, it is well known that correlations among assets increase, causing severe and widespread losses across the market. This phenomenon significantly limits investors’ ability to reduce risk by diversifying portfolio exposures across different assets. In such circumstances, since standard diversification may become ineffective, it may be useful to concentrate investments in a few assets—if such assets exist—that deviate from the behavior of the market. 
Transposing this idea in terms of local and global balance, we observe that during financial crises, the level of global balance in financial correlation networks increases, approaching its theoretical maximum (see \cite{bartesaghi2025a}). 
In this paper, we test the following intuitive hypothesis: significant deviations of a node’s local balance from the network’s global balance can help identify nodes (assets) that potentially hedge against losses during financial crises and outperform the reference market.

Although the present approach is implemented using return correlations, it overcomes some well-known limitations of standard correlation. For example, comparing a global indicator computed on the entire network with a local 
indicator associated with a single node allows us to move beyond the pairwise structure of standard correlation. Moreover, the present approach not only identifies the assets in which allocations should be concentrated but also the periods during which such concentration may be beneficial. To support this intuition, we present an empirical application using different datasets of real financial data, in both descriptive and predictive contexts. We also conduct a sensitivity analysis to determine when a deviation of local balance from global balance can be considered significant.


\section{Data, notation and preliminaries}

Given $N$ risky assets, the correlation matrix of their returns, $\bf{C}$, and the signed network $G$ derived from $\bf{C}$, the local and global balance indices are defined as follows:
\[
\kappa(G) 
	=\frac{\sum_{i=1}^{N}[e^{\bf C}]_{ii}}{\sum_{i=1}^{N}[e^{|{\bf C}|}]_{ii}}
	=\frac{\sum_{i = 1}^N e^{\lambda_i}}{\sum_{i = 1}^N e^{\overline{\lambda}_i}}, 
    \quad \quad \kappa_{i}(G):=\frac{[e^{\bf C}]_{ii}}{[e^{|{\bf C}|}]_{ii}} 
\]
  
where $\lambda_i$ and $\overline{\lambda}_i$ are, respectively, the eigenvalues of $\bf{C}$ and $|\bf{C}|$, with $i = 1,\ldots,N$, and $e^{\bf C}$ represents the matrix exponential.




The application is conducted on four datasets, each identified by the name of the corresponding financial index. The datasets contain daily log returns from 05/01/2005 to 17/09/2020. The indices considered are DAX ($N=18$), ESX ($N=42$), FTSE ($N=80$), and NIKKEI ($N=199$). 
The number $N$ of stocks in each dataset is smaller than the nominal number of index constituents, as the analysis is restricted to stocks that remained in the index throughout the reference period. The experiments are performed using a standard rolling-window setting, where the window length $\Delta T$ satisfies the condition $\Delta T \geq N$. This assumption ensures that $\bf{C}$ is a full-rank matrix. In the applications, the rolling window is shifted by a step of $\Delta t = 10$ days.



We start by examining the relationship between the average market return and the Sharpe ratio within a given time window, on the one hand, and the global balance calculated over the same period, on the other. As shown in Table \ref{table1} for all four datasets, this relationship is negative. This indicates that the values of $\kappa(G)$ increase when average returns and Sharpe ratios decrease. This is a sign that the global balance of a financial correlation network can be interpreted as a measure of systemic risk, accordingly to the hypothesis presented in \cite{bartesaghi2025a}.

\begin{center}
\begin{table}[H]
\centering{}
\begin{tabular}{lcc}
\toprule
&	Global balance - AMR & Global balance - SR  \\
\midrule
DAX  &	$-0.262$ ($-0.317$) & $-0.313$ ($-0.350$) \\
ESX  &	$-0.177$ ($-0.274$) & $-0.227$ ($-0.316$)\\
FTSE  &	$-0.215$ ($-0.324$) & $-0.199$ ($-0.325$) \\
NIKKEI  & $-0.236$ ($-0.236$) & $-0.250$ ($-0.233$)\\
\bottomrule
\hfill
\end{tabular}
\caption{Pearson correlations (Spearman correlations within brackets) between global balance and average market return (AMR) and between global balance and Sharpe ratio (SR) for the four datasets.}
\label{table1}
\end{table}
\end{center}

This initial evidence is fundamental for the implementation of the investment strategy. In general, a deviation of a node’s local balance from the global balance of the underlying network only indicates that the node behaves differently from the others, without providing information on the direction of this divergence. In contrast, when the global balance of a financial correlation signed network is high, all nodes (assets) tend to report losses. Therefore, assets with a local balance that deviates from the global one are those that outperform the rest of the market.




\section{Numerical analysis}

The idea of concentrating the portfolio in a few assets that locally deviate from the systemic behavior of the market, in time windows characterized by high global balance, is implemented in the following experiment.
We select two thresholds, $\tau_G$ and $\tau_L$, to simultaneously identify when the global balance is high and the local balance deviates significantly from it. 
Specifically, the set of selected assets is $S_A=\{i=1, \ldots N:\kappa(G) \geq \tau_G, \kappa_G(t) - \kappa_i(t) \geq \tau_L\}$, whose cardinality $|S_A|$ gives the number of selected assets.

Since the specific allocation rule is not the focus of the present work, we invest in an equally weighted portfolio of the selected assets when the conditions are met, and in an equally weighted portfolio ($1/N$) otherwise.
Finally, the performance of the $1/N$ portfolio over the entire period, measured in terms of return and Sharpe ratio, is compared with that of the proposed strategy.




\subsection{In sample results}

We begin by analyzing the in-sample performance, based on the average returns of the last five days of each examined window. Figure \ref{fig1} presents the value -- in terms of price -- of the proposed portfolio strategy (red line) compared with that of the $1/N$ portfolio (blue line), for all datasets, with different values of the thresholds $\tau_L$ and $\tau_G$. What clearly emerges is that the {proposed portfolio outperforms the $1/N$ portfolio across all datasets. Moreover, the higher final value does not imply higher volatility for the improved strategy.

\begin{figure}[H]
	\centering
	\subfloat[]{\includegraphics[width=0.45\textwidth]{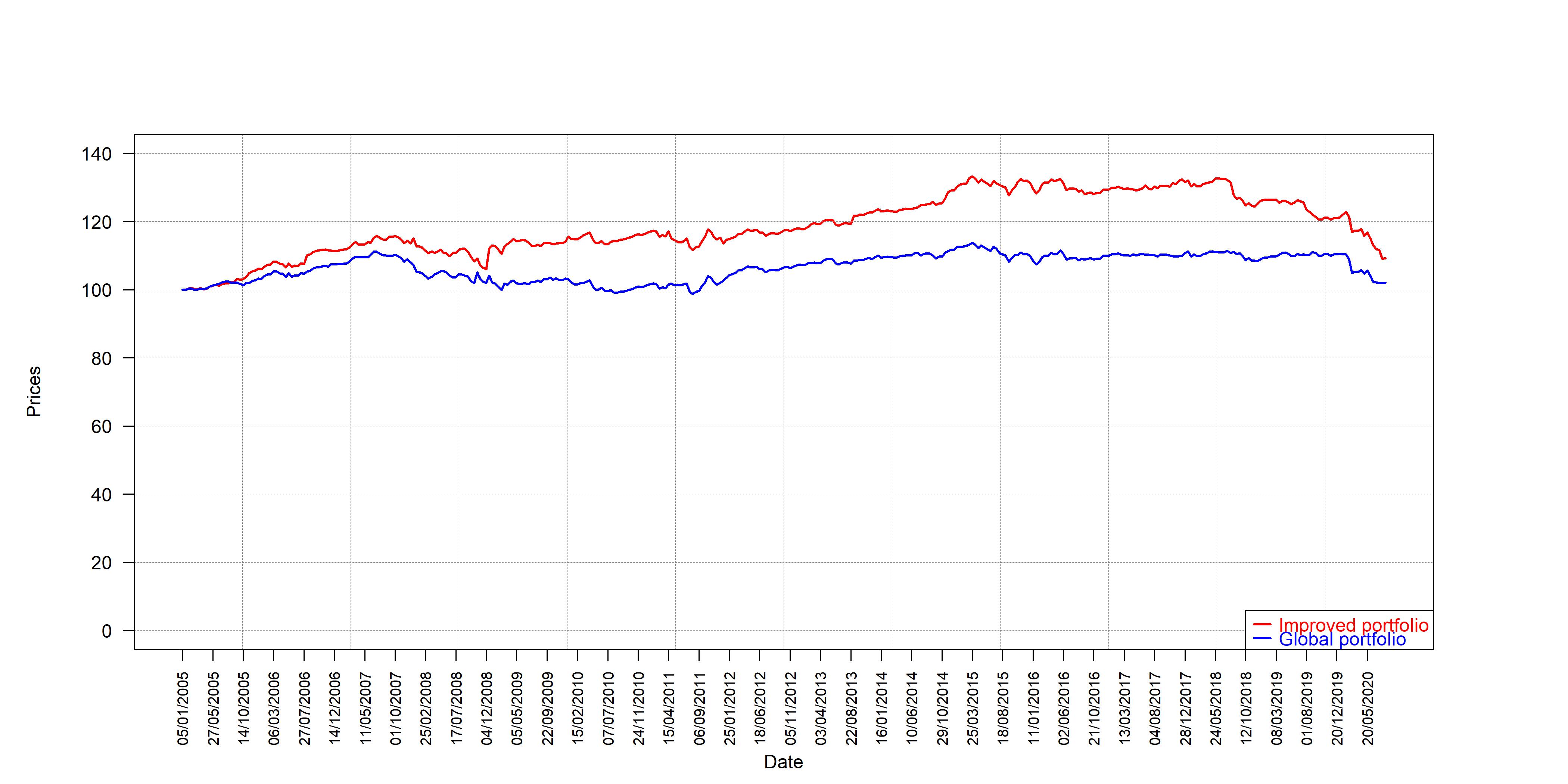}}
	\subfloat[]{\includegraphics[width=0.45\textwidth]{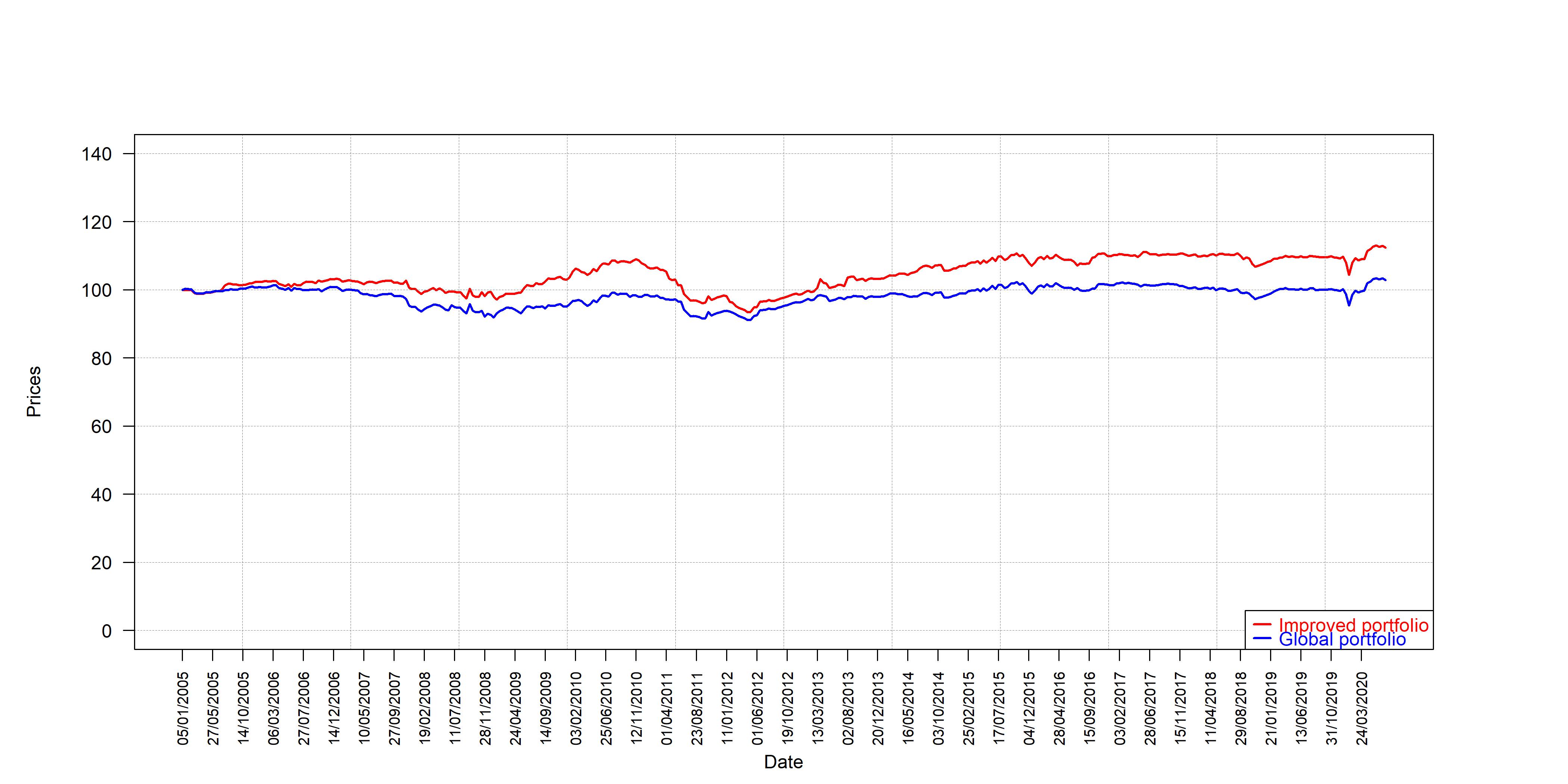}}\\
	\subfloat[]{\includegraphics[width=0.45\textwidth]{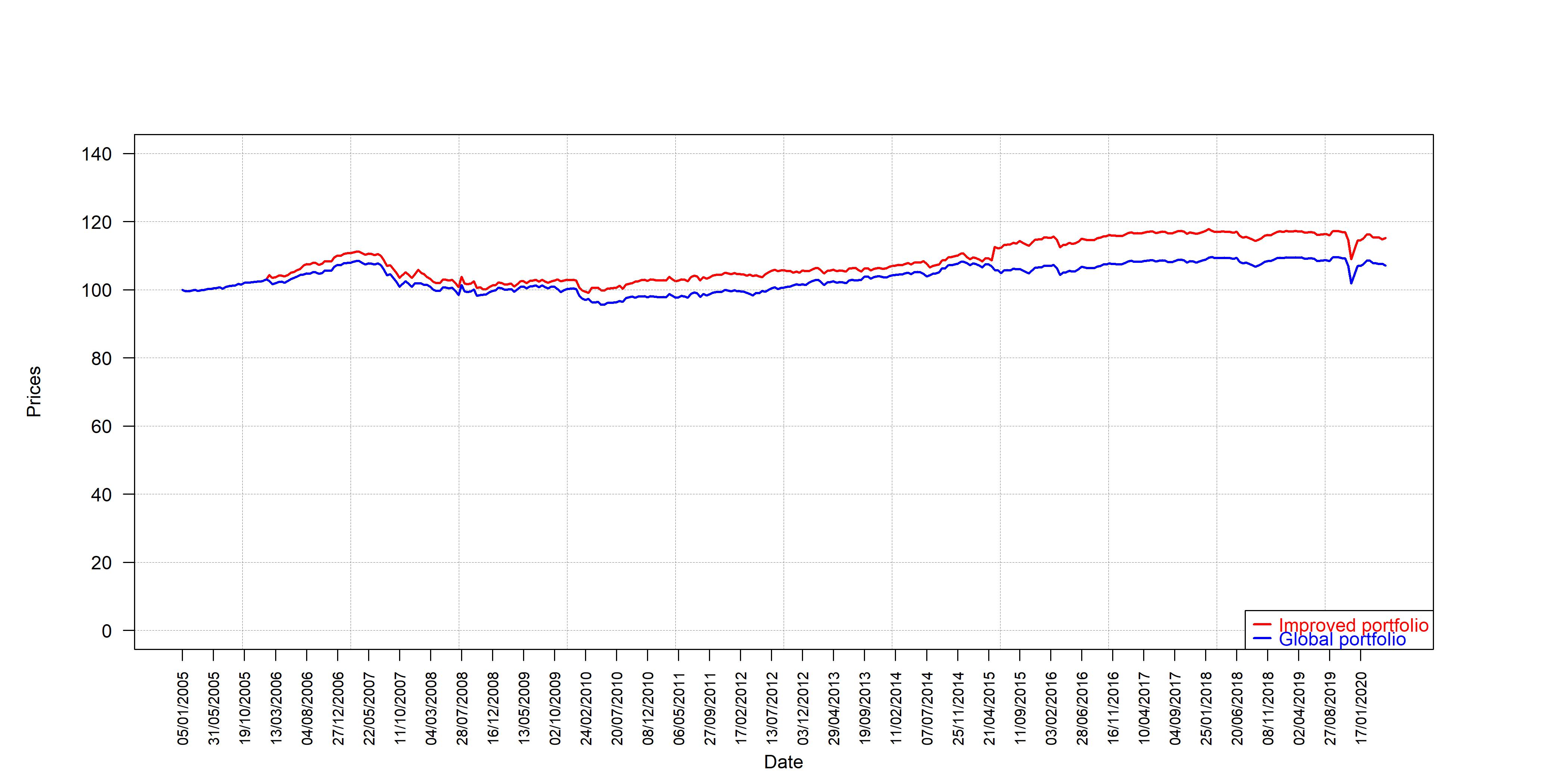}}
	\subfloat[]{\includegraphics[width=0.45\textwidth]{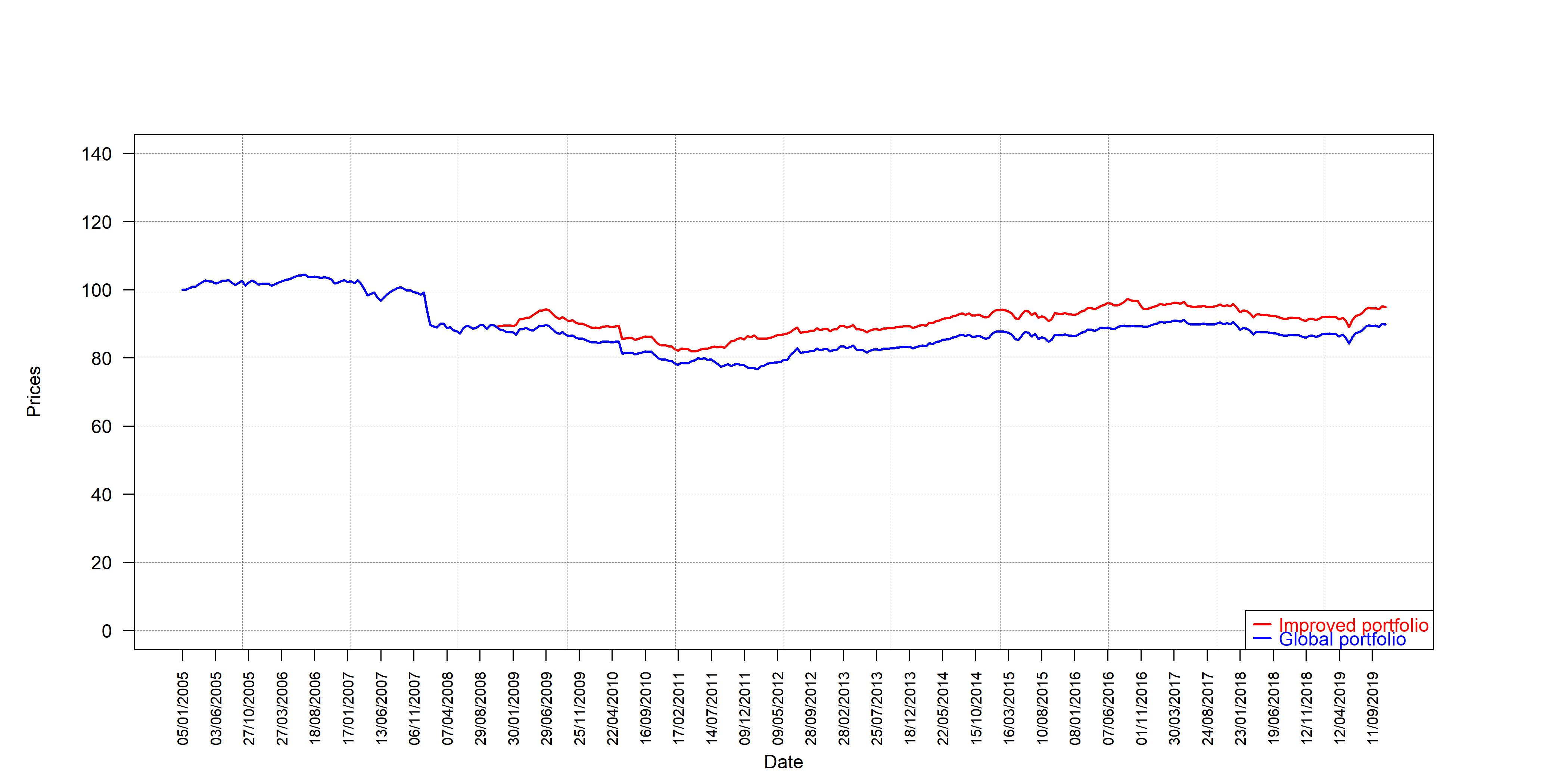}}
	\caption{Prices of the standard portfolio (blue line) and the improved portfolio (red line): 
    (a) DAX, $\tau_{G}=0.75$ and $\tau_{L}=0.35$; (b) ESX, $\tau_{G}=0.75$ and $\tau_{L}=0.25$; (c) FTSE, $\tau_{G}=0.90$ and $\tau_{L}=0.40$; (d) NIKKEI, $\tau_{G}=0.90$ and $\tau_{L}=0.45$;}
	\label{fig1}
\end{figure}

The chosen thresholds are different for each examined dataset. An
appropriate calibration of these values is required for the improved portfolio to outperform the baseline portfolio. The robustness of these results with respect to the chosen parameters is discussed in Section \ref{robustness}, where we present a dedicated sensitivity analysis.

To further support the results, we analyze the probability distributions of returns and corresponding Sharpe ratios of the selected stocks within each time window and compare them with those of the remaining stocks. To ensure consistent comparisons between distributions and to avoid biases related to the smaller number of selected assets relative to the remaining ones, we proceed as follows.
We randomly sample from the remaining stocks a number of securities equal to $|S_A|$.
This random selection is repeated 10,000 times, and the results are averaged to obtain statistics that enable a robust comparison.

In Figure \ref{fig2}, we present the density functions and box plots for the NIKKEI dataset. As in Figure \ref{fig1}, the red line represents the selected stocks, while the blue line represents the others. Both the distributions and the box plots show that the distribution of the selected stocks is characterized by a higher mean, comparable variance, a thinner left tail, a fatter right tail, and greater skewness. All these findings are visually supported by Figure \ref{fig2} and confirmed by the values of the distribution moments, reported in the Table \ref{table2} together with the threshold values.



\begin{figure}[H]
	\centering
	\subfloat[]{\includegraphics[width=0.45\textwidth]{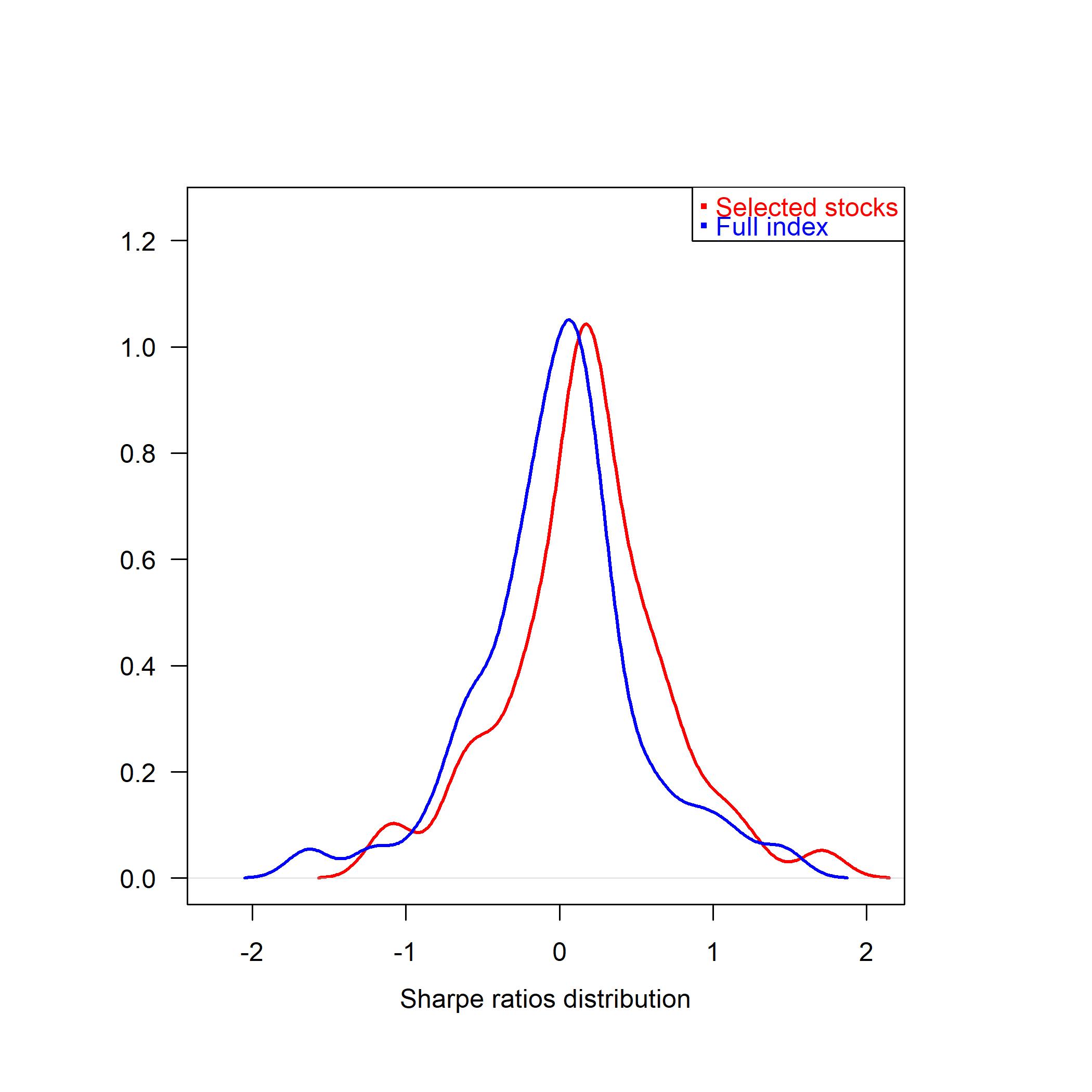}}
	\subfloat[]{\includegraphics[width=0.45\textwidth]{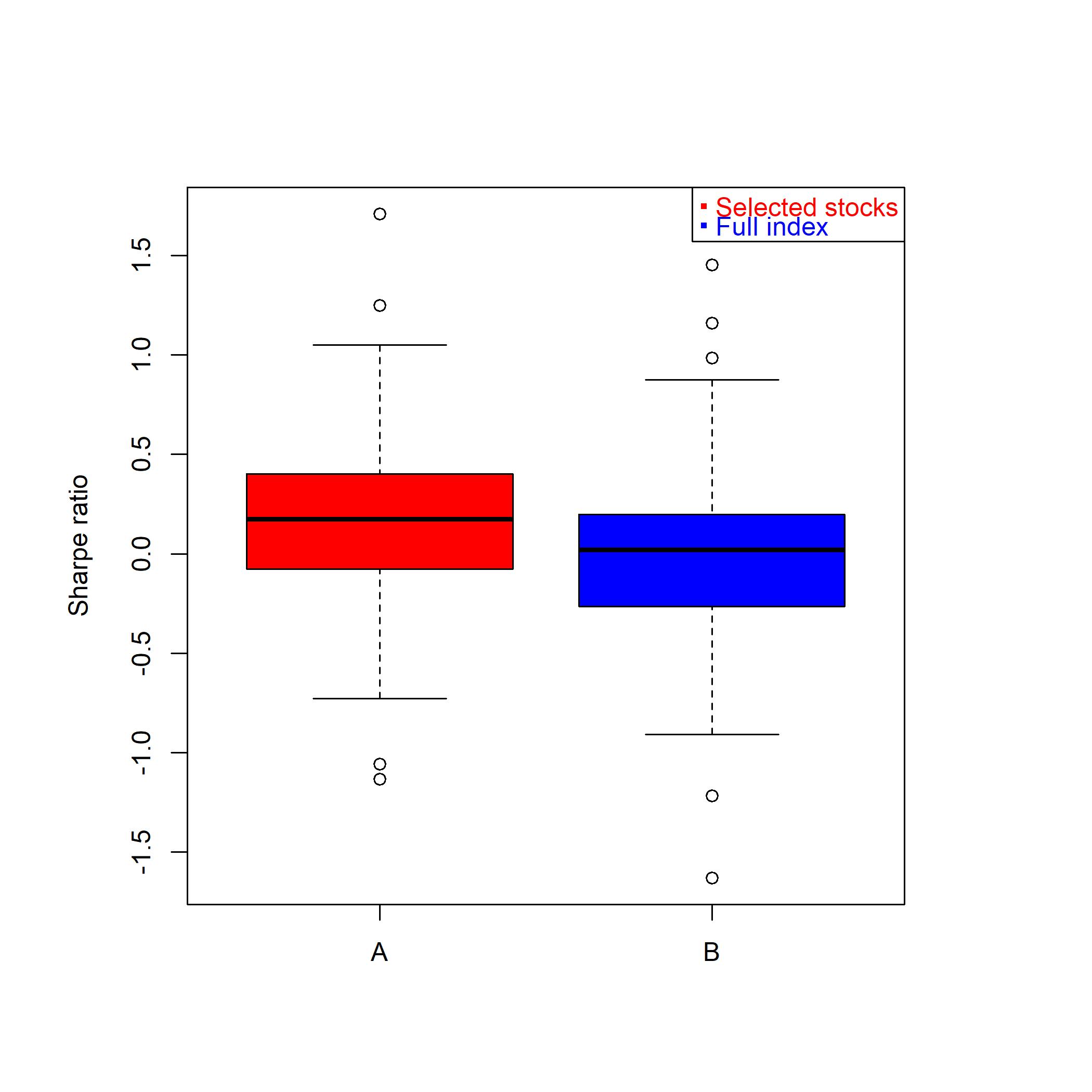}}
	\caption{In sample densities, panel (a) and box plot, panel (b), for the Sharpe ratios of the selected stocks (in red) and the remainder of the dataset (in blue) for the NIKKEI index.}
	\label{fig2}
\end{figure}

\begin{table}[h!]
\centering
\caption{In-sample distribution moments for Sharpe ratios of selected stocks and remaining dataset for NIKKEI index.}
\label{tab:sharpe_ratios_nikkei_insample}
\begin{tabular}{lcccccc}
\toprule
\textbf{Group} & \boldmath$\tau_G$ & \boldmath$\tau_L$ & \textbf{Mean} & \textbf{Variance} & \textbf{Skewness} & \textbf{Kurtosis} \\
\midrule
Selected stocks & 0.90 & 0.46 & 0.1697782 & 0.2853324 & 0.06643058 & 0.6962828 \\
Remaining stocks & 0.90 & 0.46 & -0.01147054 & 0.2928941 & -0.09771086 & 1.240045 \\
\bottomrule
\end{tabular}
\label{table2}
\end{table}

\subsection{Sensitivity analysis with respect to the choice of the thresholds} \label{robustness}

The robustness of the in-sample results is tested on the NIKKEI dataset, which has the largest number of assets ($N=199$). Specifically, we conduct a sensitivity analysis on the two parameters, $\tau_G$ and $\tau_L$. Figure \ref{fig3} shows how the distributions of the Sharpe ratio evolve as $\tau_G$ increases from $0.75$ (panel (a)) to $0.90$ (panel (d)), in increments of $0.05$, while $\tau_L$ remains constant at $\tau_L = 0.45$.
It is evident how the distribution of the selected securities consistently outperforms that of randomly selected securities within the global portfolio. These results confirm that the findings are independent of the specific value of $\tau_G$, which can therefore be used to calibrate the desired overall risk level and determine the number of time windows in which the improved portfolio should replace the baseline one. In general, when both the global and local balance thresholds are reduced, the two distributions converge, as gradually all the securities in the dataset are selected. Table \ref{table3} shows the values of the moments of the two distributions for the four values of $\tau_G$.

\begin{figure}[H]
	\centering
	\subfloat[]{\includegraphics[width=0.25\textwidth]{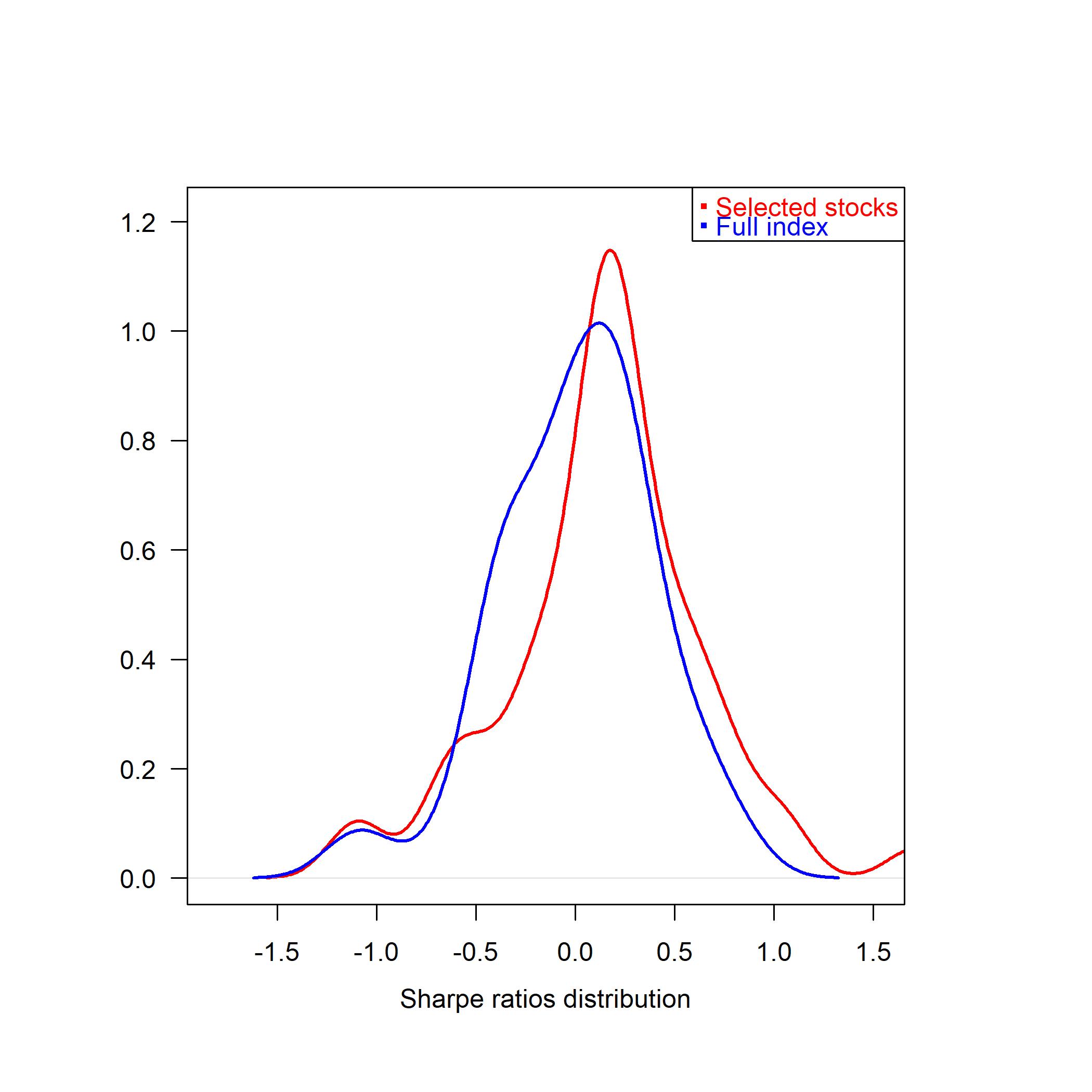}}
	\subfloat[]{\includegraphics[width=0.25\textwidth]{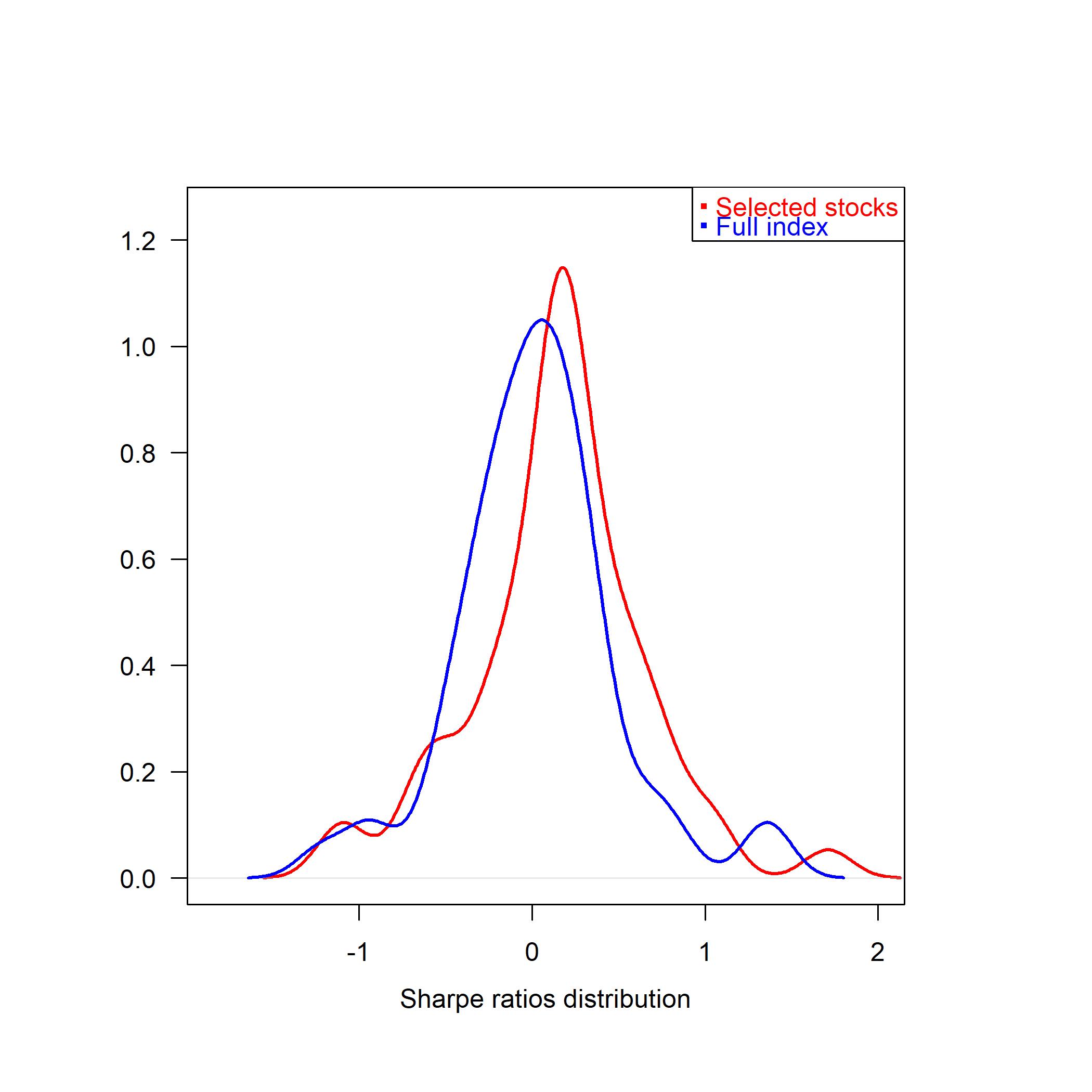}}
	\subfloat[]{\includegraphics[width=0.25\textwidth]{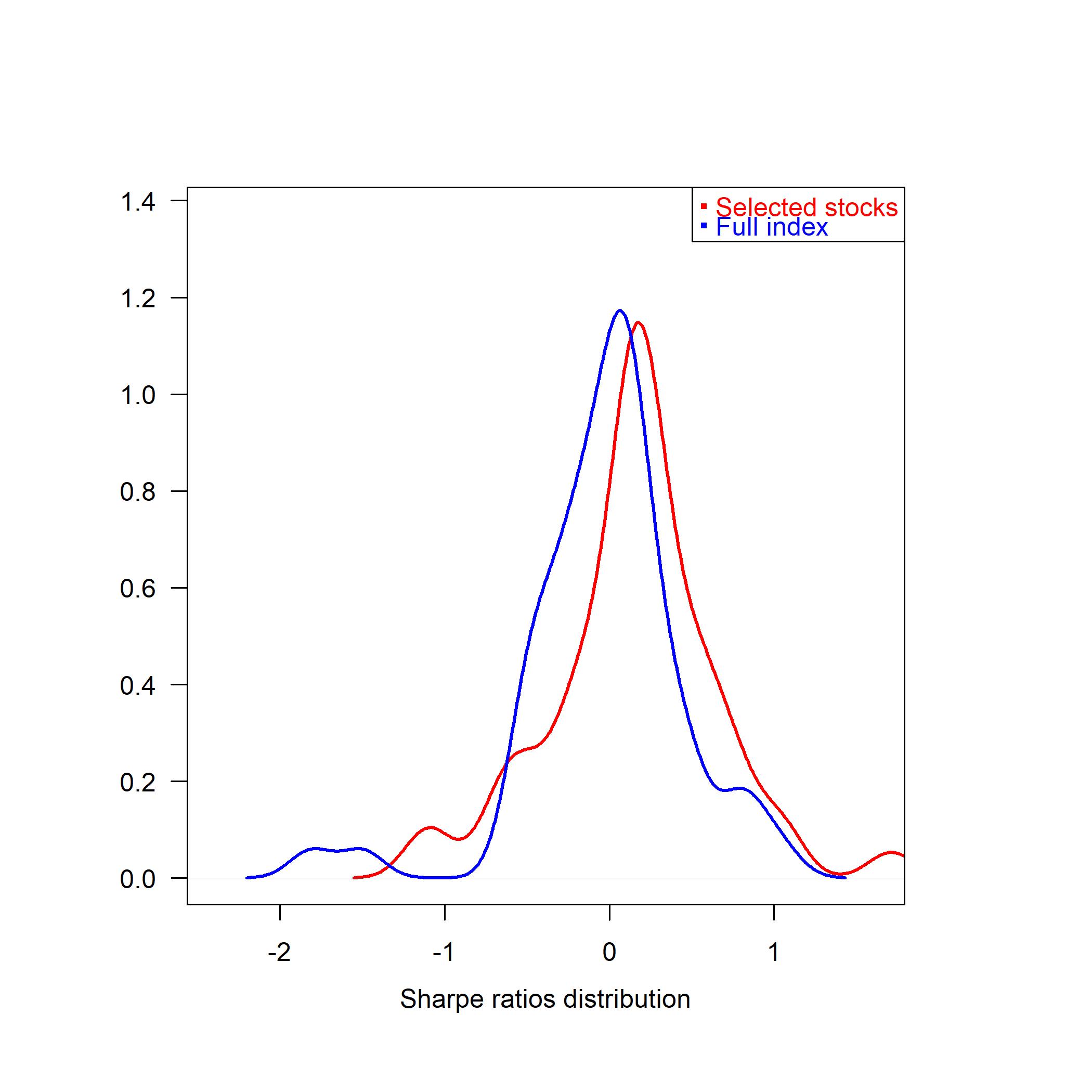}}
	\subfloat[]{\includegraphics[width=0.25\textwidth]{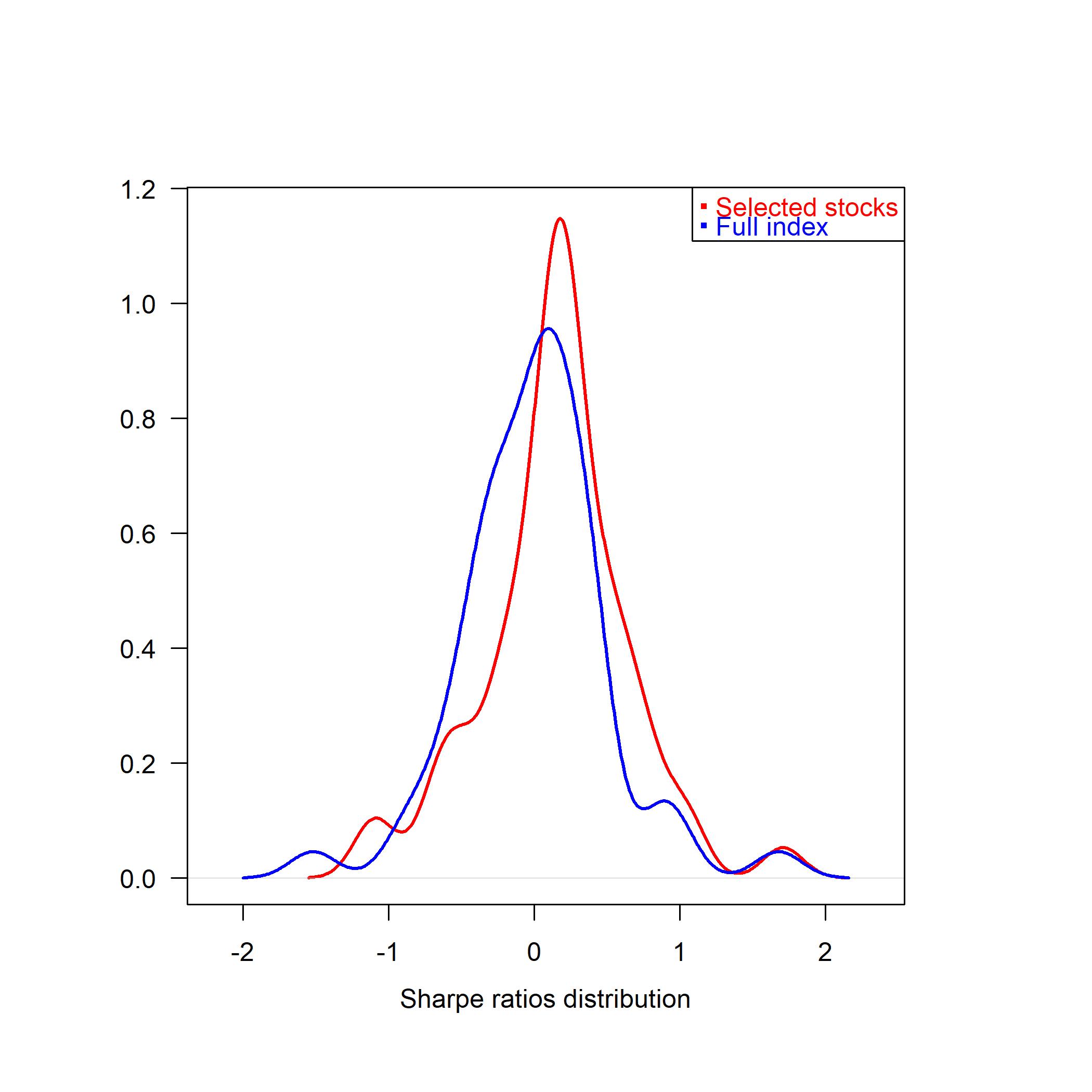}}\\
	\subfloat[]{\includegraphics[width=0.25\textwidth]{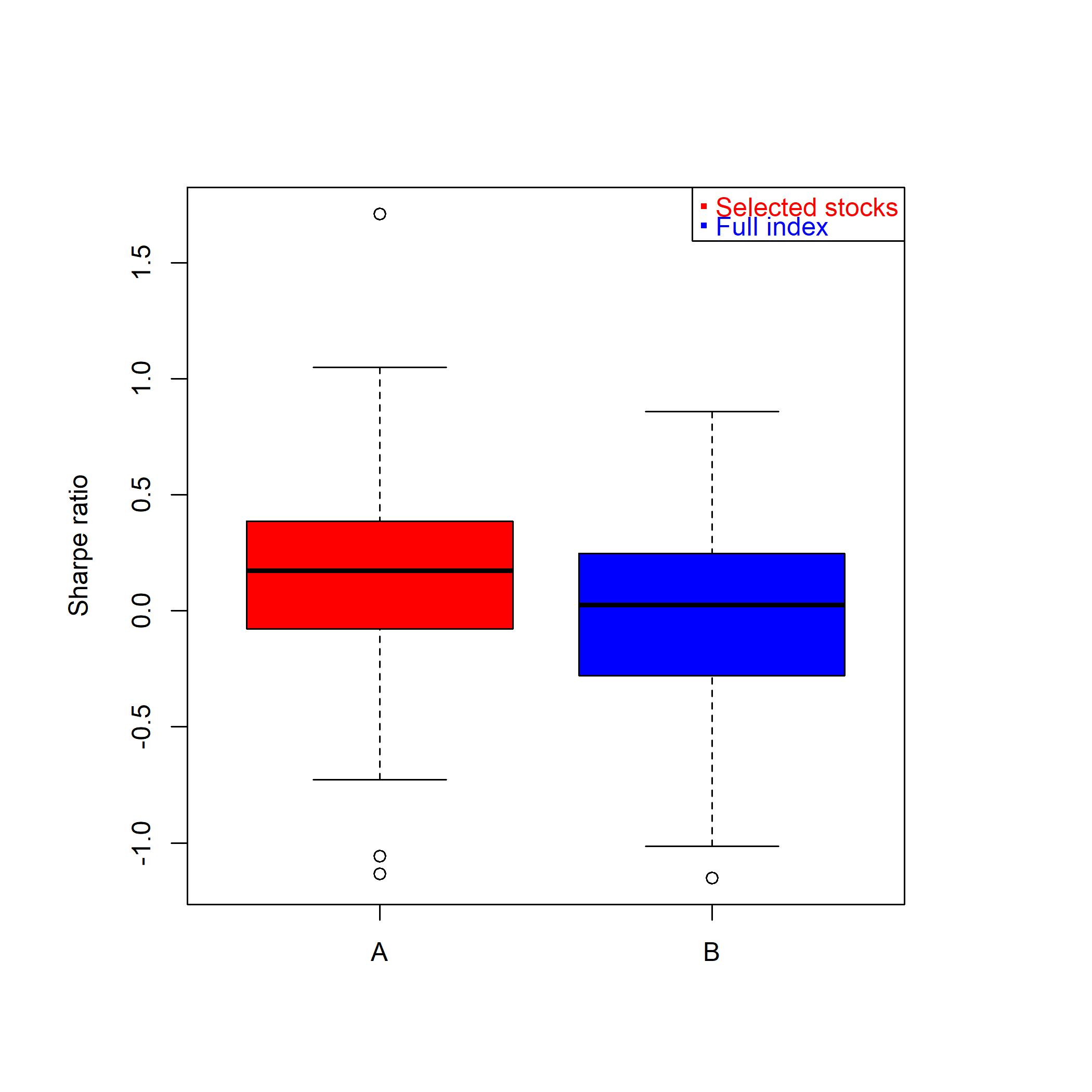}}
	\subfloat[]{\includegraphics[width=0.25\textwidth]{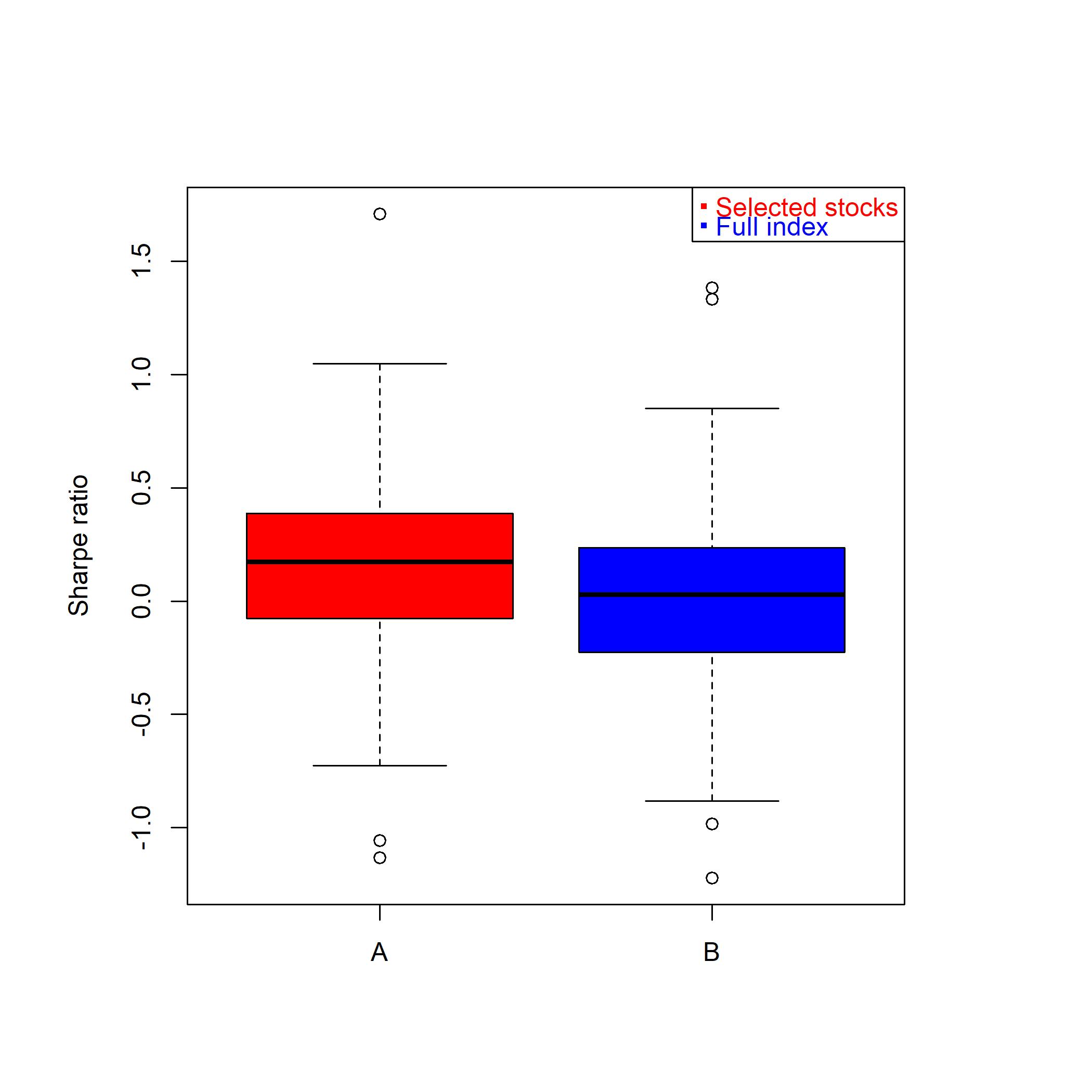}}
	\subfloat[]{\includegraphics[width=0.25\textwidth]{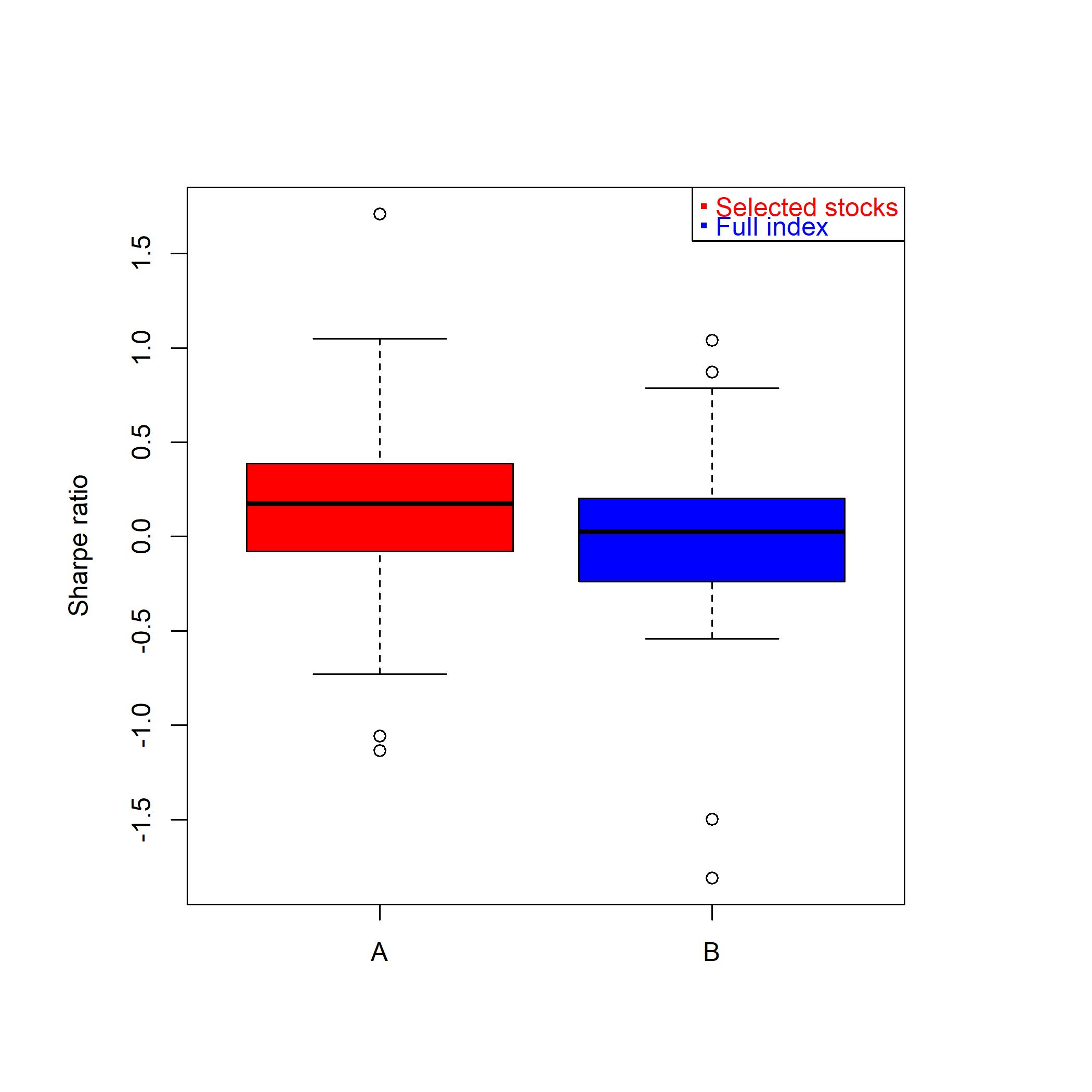}}
	\subfloat[]{\includegraphics[width=0.25\textwidth]{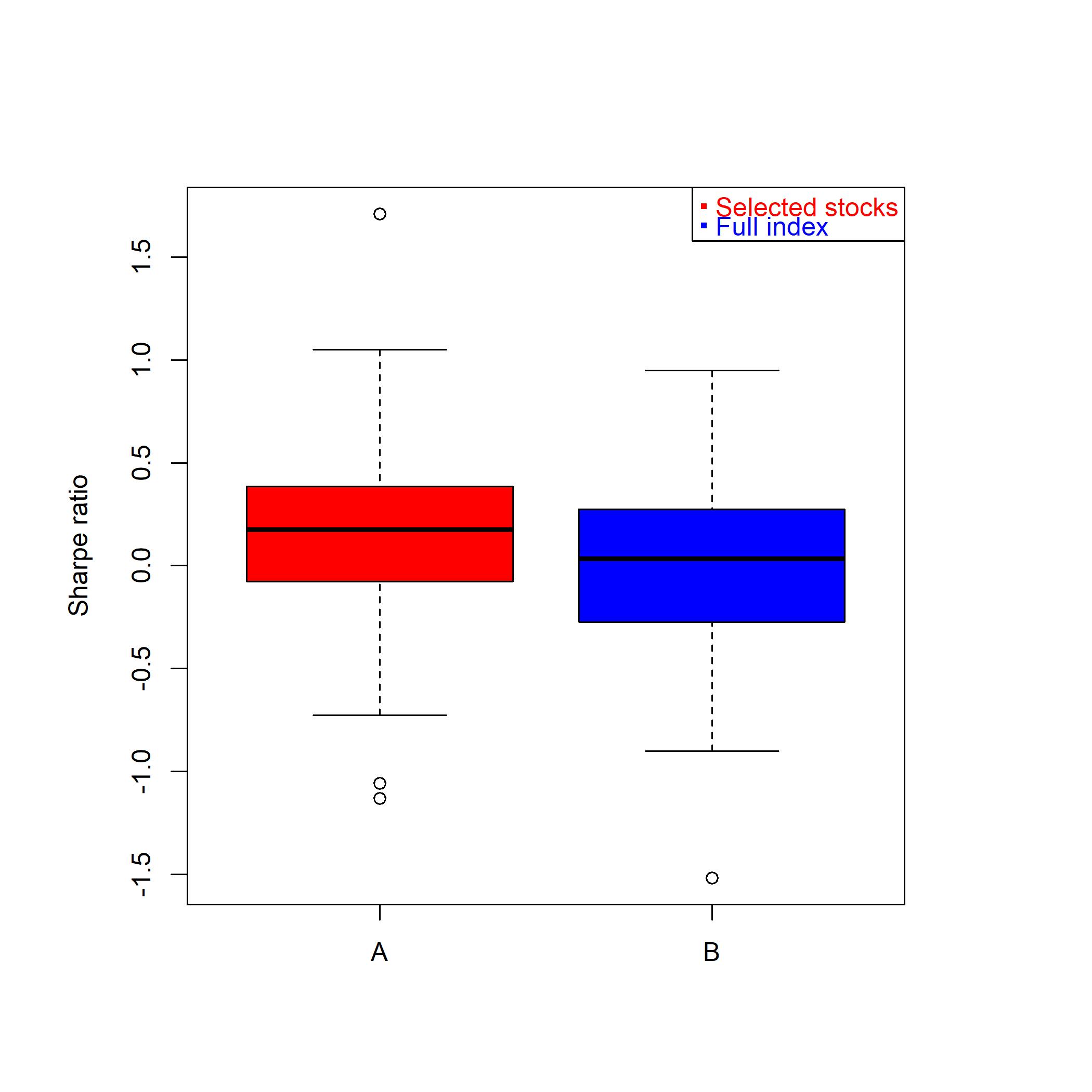}}
    \caption{Densities for the Sharpe ratios of two groups of selected stocks (in red) and the remainder of the dataset (in blue) for NIKKEI index; $\tau_{L}=0.45$ for all plots.}
	\label{fig3}
\end{figure}

\begin{table}[h!]
\centering
\caption{Summary statistics for Sharpe ratios of selected stocks and global dataset  across different threshold configurations for NIKKEI index.}
\label{sharpe_ratios_nikkei}
\begin{tabular}{ccccccc}
\toprule
\textbf{Panels} & \boldmath$\tau_G$ & \textbf{Group} & \textbf{Mean} & \textbf{Variance} & \textbf{Skewness} & \textbf{Kurtosis} \\
\midrule
(a), (e) & 0.75 & Selected & 0.1507192 & 0.2576365 & 0.008535879 & 1.064017 \\
& & Global & 0.007813921 & 0.1589361 & -0.4683775 & 0.4643931 \\
\midrule
(b), (f) & 0.80 & Selected & 0.1507192 & 0.2576365 & 0.008535879 & 0.6962828 \\
& & Global & 0.02143808 & 0.2239296 & 0.3006724 & 1.408511 \\
\midrule
(c), (g) & 0.85 & Selected & 0.1507192 & 0.2576365 & 0.008535879 & 1.064017 \\
& & Global & -0.02057534 & 0.2327003 & -1.020191 & 3.301265 \\
\midrule
(d), (h) & 0.90 & Selected & 0.1507192 & 0.2576365 & 0.008535879 & 1.064017 \\
& & Global & 0.002948375 & 0.2498295 & 0.2155788 & 2.109817 \\
\bottomrule
\end{tabular}
\label{table3}
\end{table}

A higher sensitivity is observed for $\tau_L$, which affects the number of selected securities. As the gap $\kappa_G(t) - \kappa_i(t)$ increases, the cardinality $|S_A|$ decreases accordingly, and the resulting distribution becomes less meaningful. Therefore, an optimal threshold value for the gap exists and could be found by maximizing the difference in Sharpe ratios, possibly without excessively reducing the portfolio size. Figure \ref{fig4}, where $\tau_G = 0.90$ and $\tau_L$ varies from $0.44$ to $0.50$, shows that maintaining $\tau_L$ within a relatively narrow range yields robust results, confirming the existence of a profitable region for applying the strategy. Table \ref{table4} collects the values of the moments of the two distributions for the four values of $\tau_L$.

\begin{figure}[H]
	\centering
	\subfloat[]{\includegraphics[width=0.25\textwidth]{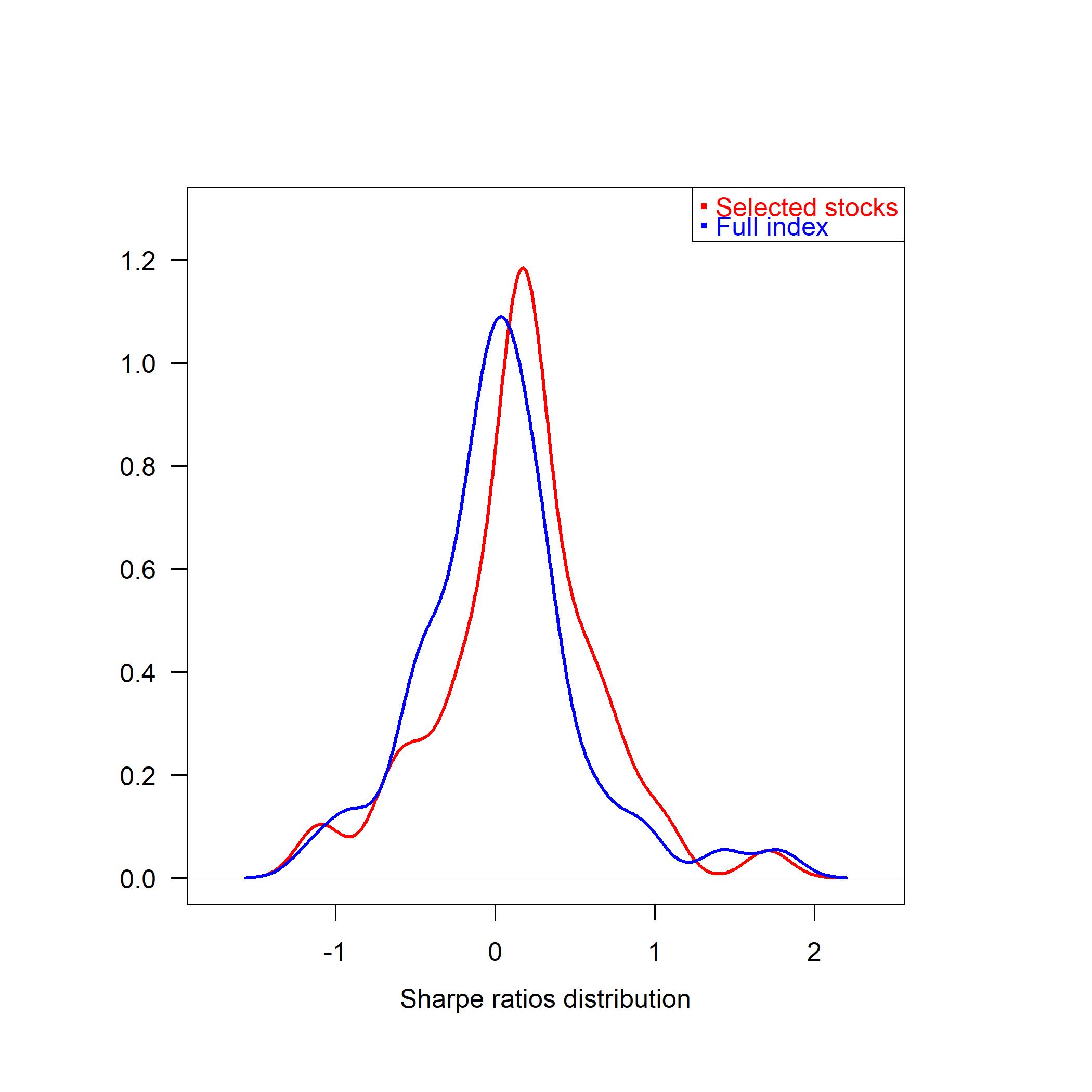}}
	\subfloat[]{\includegraphics[width=0.25\textwidth]{Final_Density_sharpe_ratio_equal_length_NIKKEI_200_10_0.9_0.46_-5}}
	\subfloat[]{\includegraphics[width=0.25\textwidth]{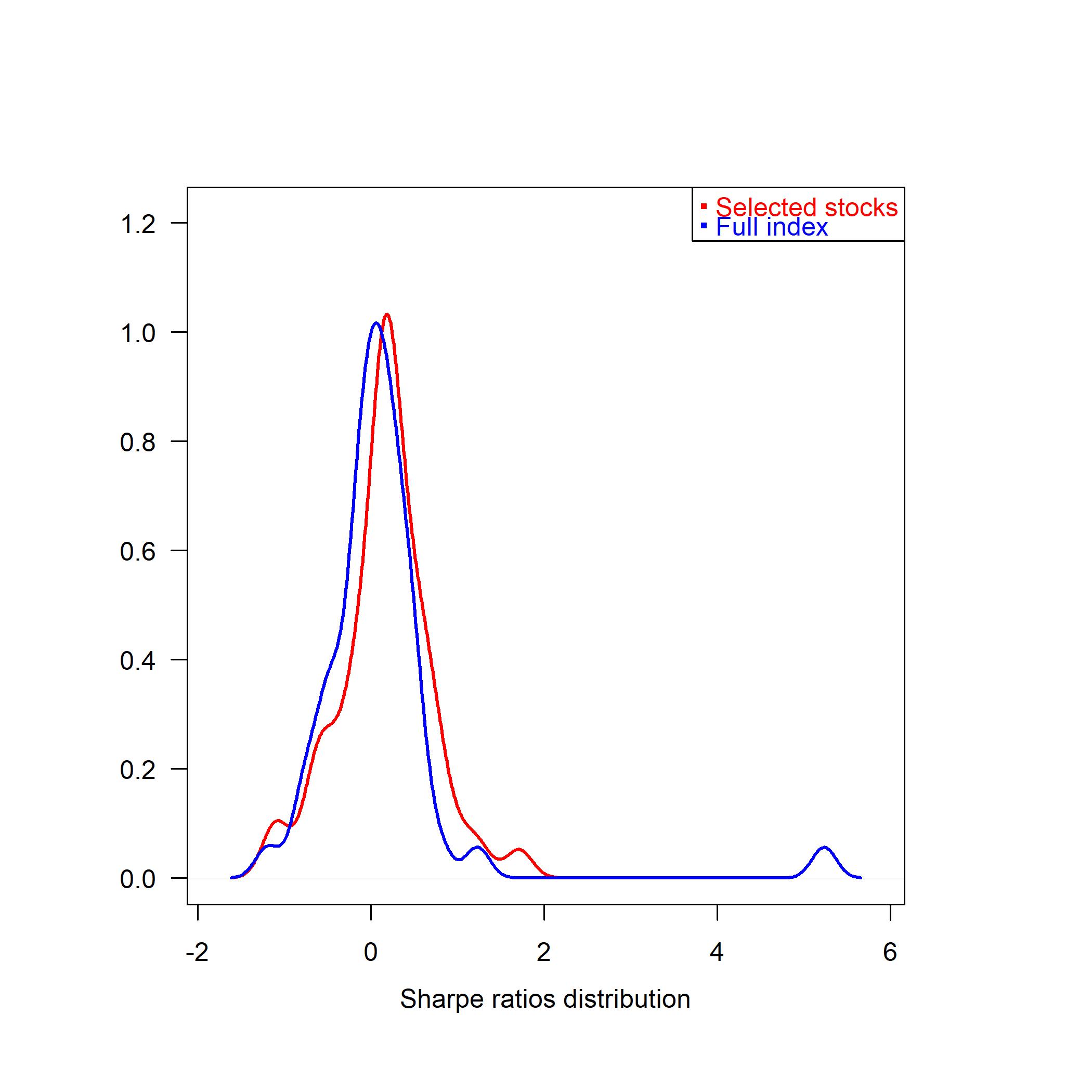}}
    \subfloat[]{\includegraphics[width=0.25\textwidth]{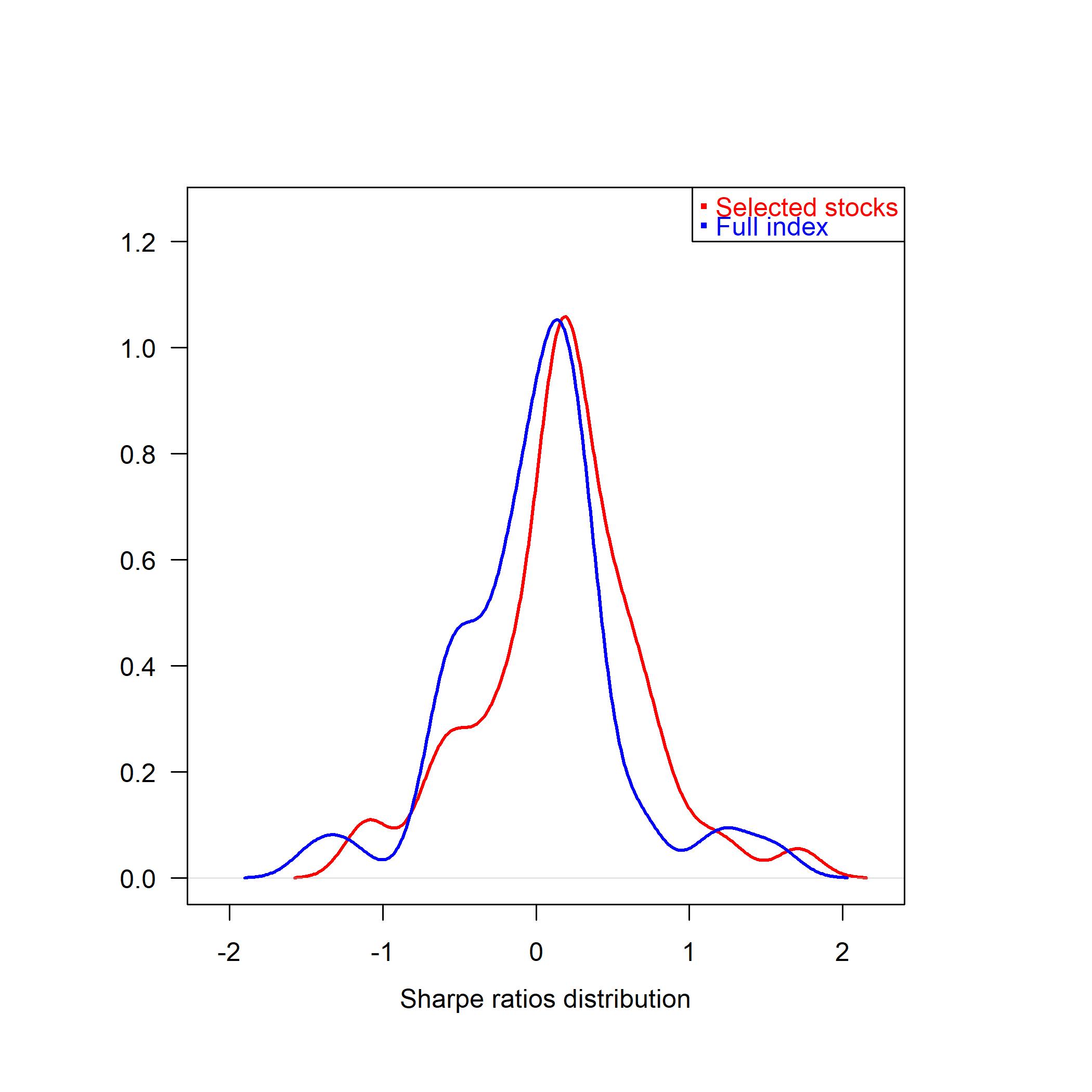}}\\
	\subfloat[]{\includegraphics[width=0.25\textwidth]{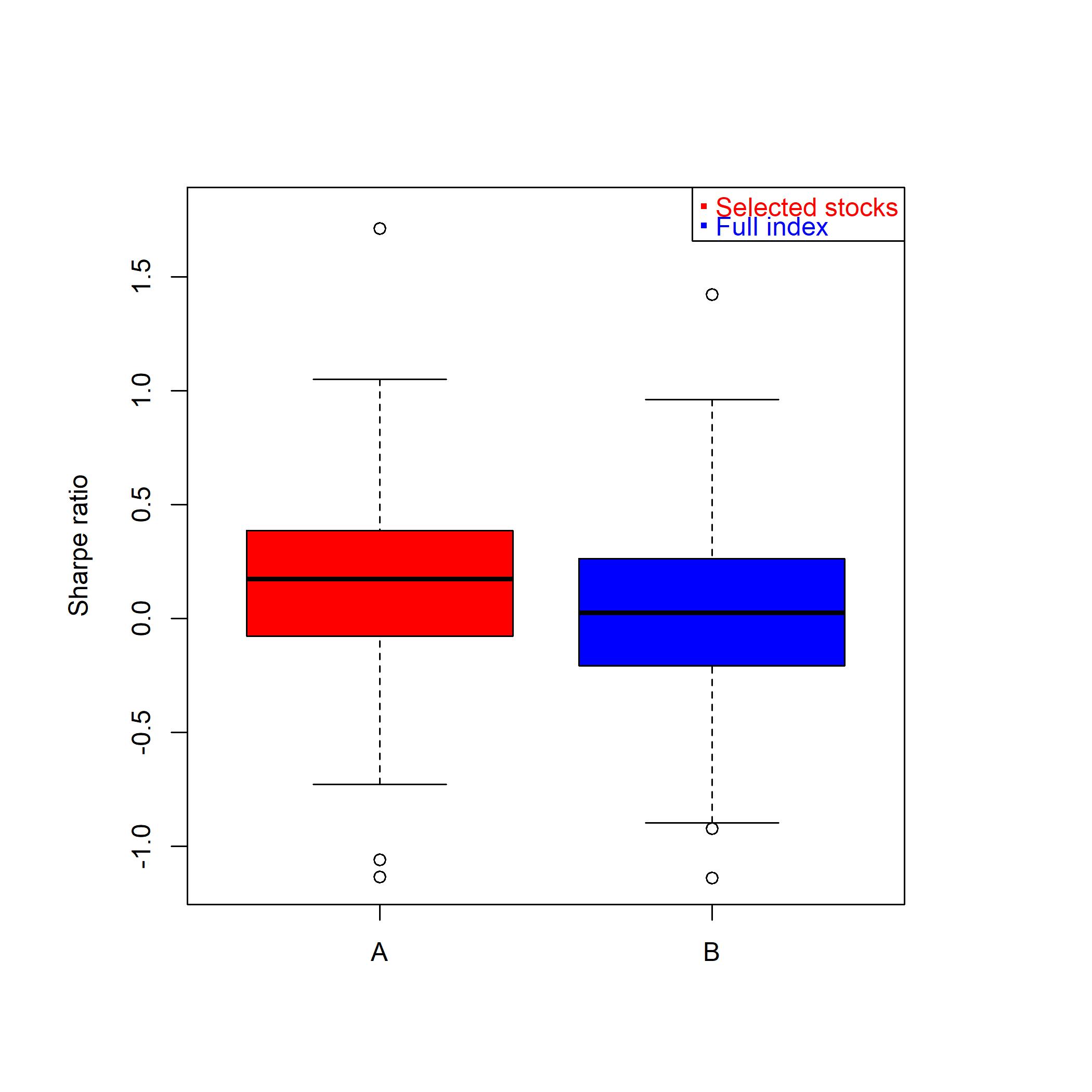}}
	\subfloat[]{\includegraphics[width=0.25\textwidth]{Final_Boxplot_sharpes_ratio_equal_length_NIKKEI_200_10_0.9_0.46_-5}}
	\subfloat[]{\includegraphics[width=0.25\textwidth]{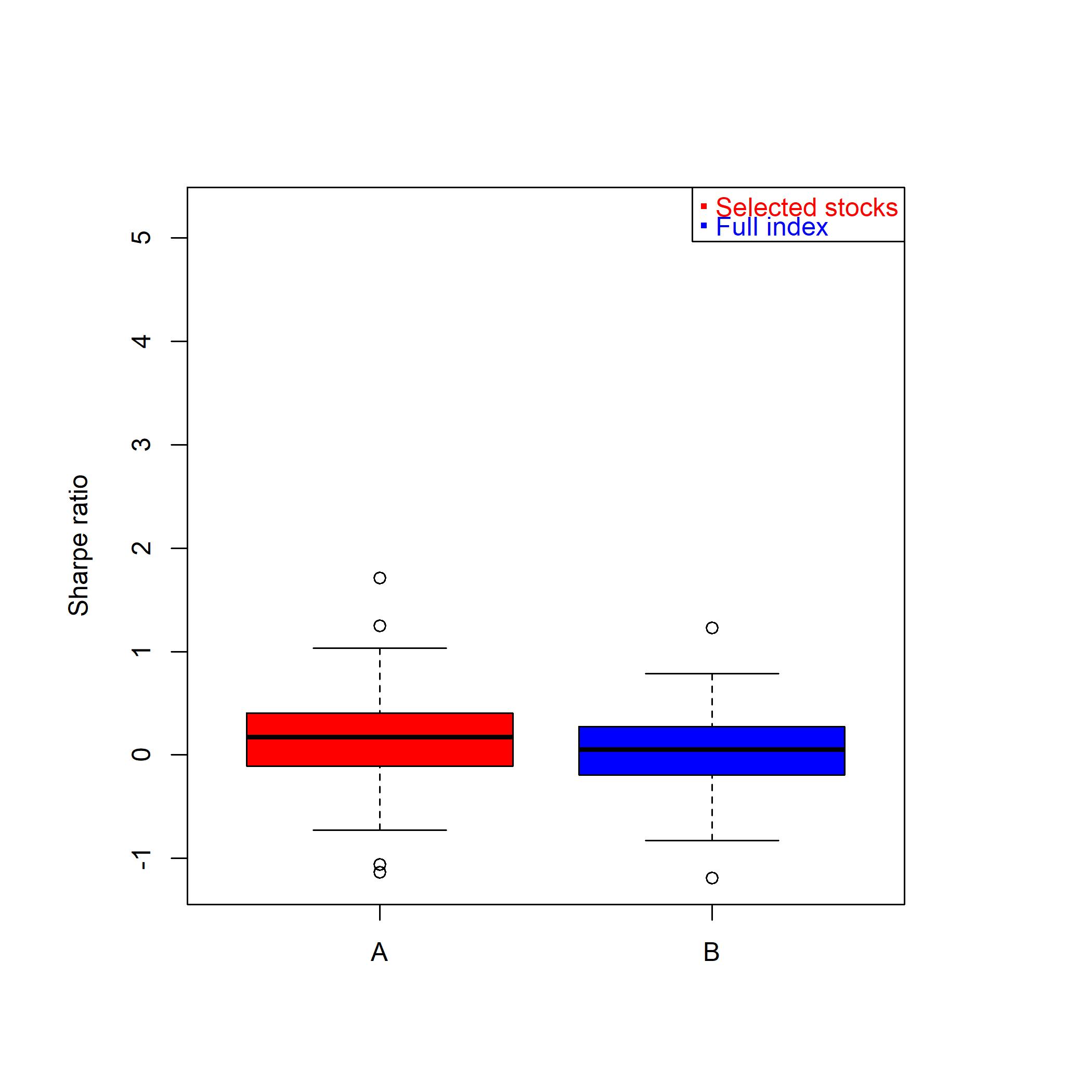}}
    \subfloat[]{\includegraphics[width=0.25\textwidth]{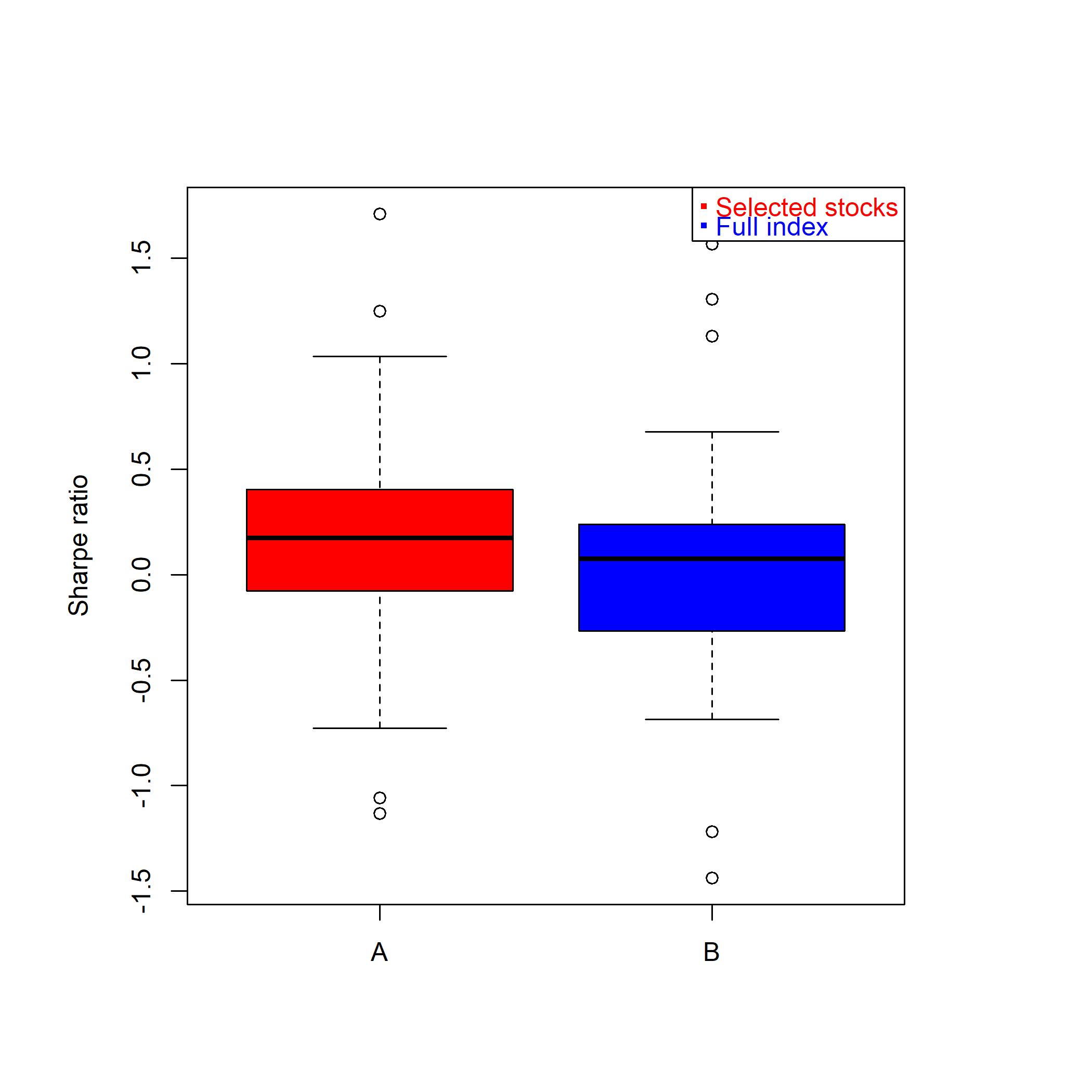}}
	\caption{Densities for the Sharpe ratios of 
    the selected stocks (in red) and the remainder of the dataset (in blue) for NIKKEI index; $\tau_{G}=0.90$ for all plots.}
	\label{fig4}
\end{figure}

\begin{table}[h!]
\centering
\caption{Summary statistics for Sharpe ratios of selected stocks and global dataset  across different threshold configurations for NIKKEI index.}
\label{tab:sharpe_ratios_nikkei_tau_l}
\begin{tabular}{ccccccc}
\toprule
\textbf{Panels} & \boldmath$\tau_L$ & \textbf{Group} & \textbf{Mean} & \textbf{Variance} & \textbf{Skewness} & \textbf{Kurtosis} \\
\midrule
(a), (e) & 0.44 & Selected & 0.1476046 & 0.2567015 & 0.02498729 & 1.093287 \\
& & Global & 0.03372156 & 0.2652361 & 0.7340246 & 1.92165 \\
\midrule
(b), (f) & 0.46 & Selected & 0.1697782 & 0.2853324 & 0.06643058 & 0.6962828 \\
& & Global & -0.01147054 & 0.2928941 & -0.09771086 & 1.240045 \\
\midrule
(c), (g) & 0.48 & Selected & 0.1607774 & 0.2839912 & 0.03776458 & 5.63365 \\
& & Global & 0.11454 & 0.7331053 & 4.157748 & 22.83232 \\
\midrule
(d), (h) & 0.50 & Selected & 0.1657216 & 0.2885198 & 0.01298107 & 0.291521 \\
& & Global & 0.01242995 & 0.1607477 & 0.2007902 & 1.410646 \\
\bottomrule
\end{tabular}
\label{table4}
\end{table}

\subsection{Out-of-sample analysis}

We extend the analysis to the out-of-sample context to evaluate whether the proposed strategy outperforms the benchmark $1/N$ in practice. Specifically, returns are averaged over the five days following the assessment window. Accordingly, we concentrate the portfolio in a few stocks based on a large balance gap (local versus global) and compare the out-of-sample Sharpe ratios of the selected stocks with those of the remaining ones. The strategy of randomly selecting from the remaining stocks a number equal to that of the selected ones is again employed to compare the corresponding distributions.

We report results only for the NIKKEI index. Since the highest sensitivity is observed for the $\tau_L$ values, we present graphs for a fixed $\tau_G$ and four different levels of $\tau_L$. Figure \ref{fig5} shows the distributions and box plots of the two groups of stocks, and Table \ref{table5} collects the values for $\tau_L$ and the corresponding  moments.

\begin{figure}[H]
	\centering
	\subfloat[]{\includegraphics[width=0.25\textwidth]{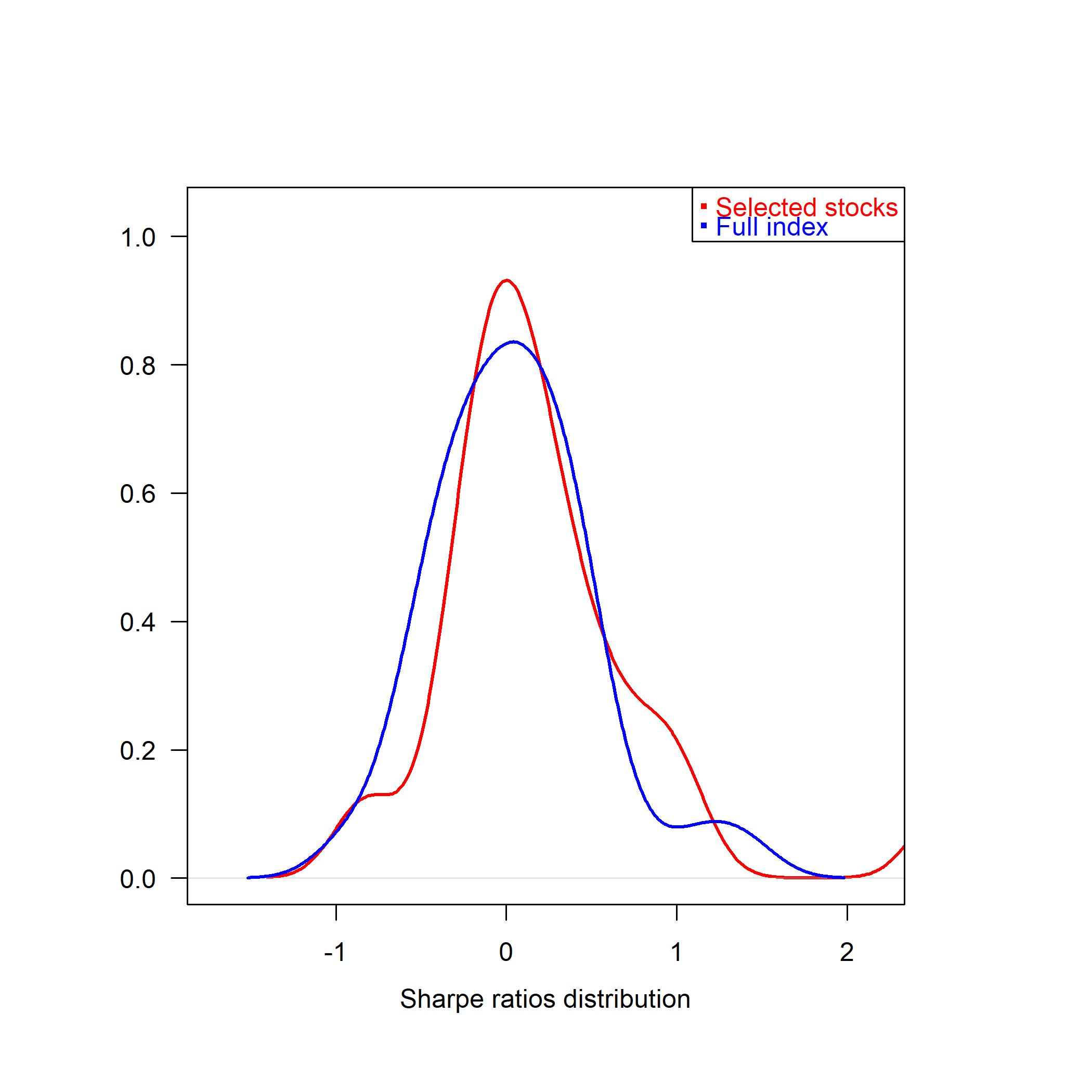}}
    \subfloat[]{\includegraphics[width=0.25\textwidth]{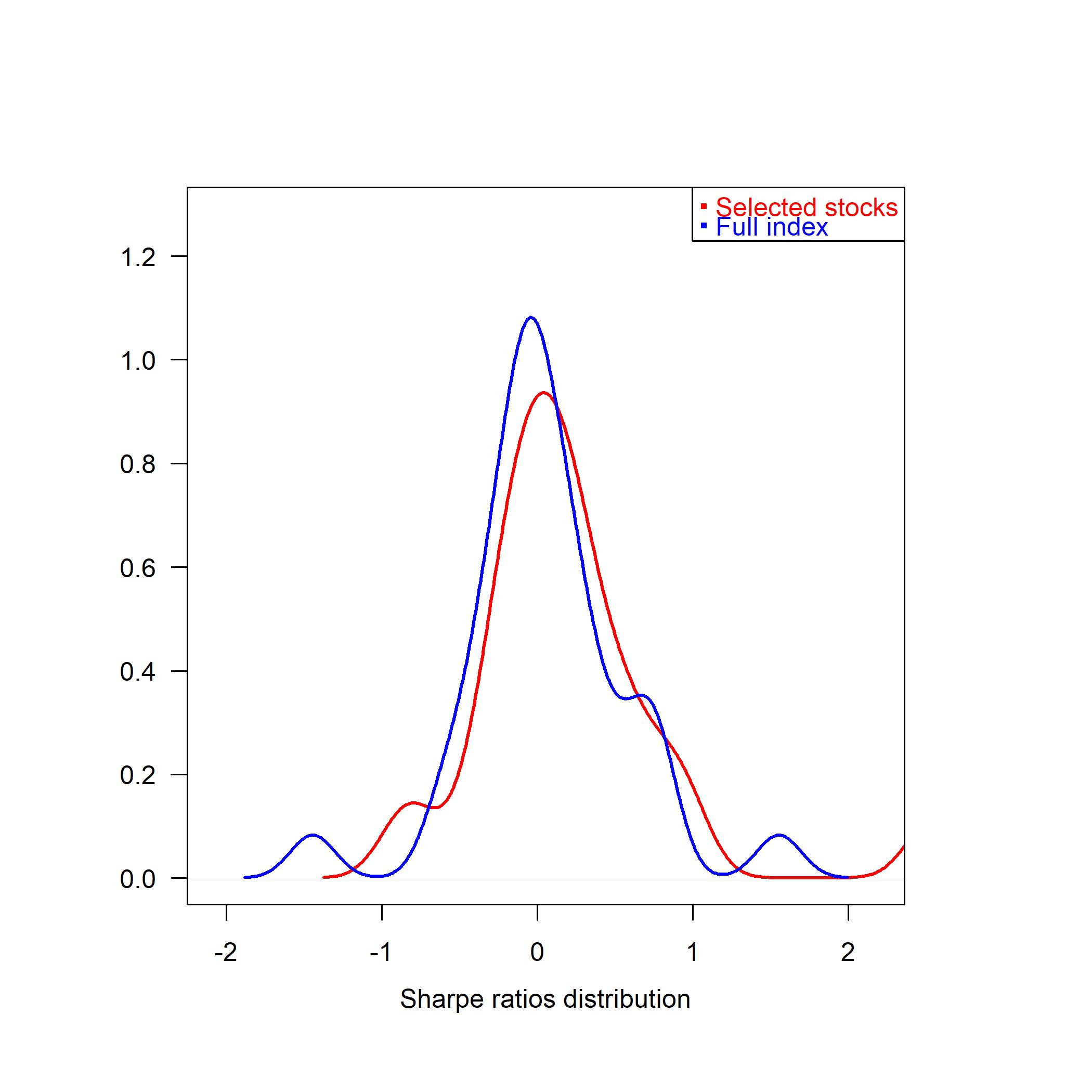}}
    \subfloat[]{\includegraphics[width=0.25\textwidth]{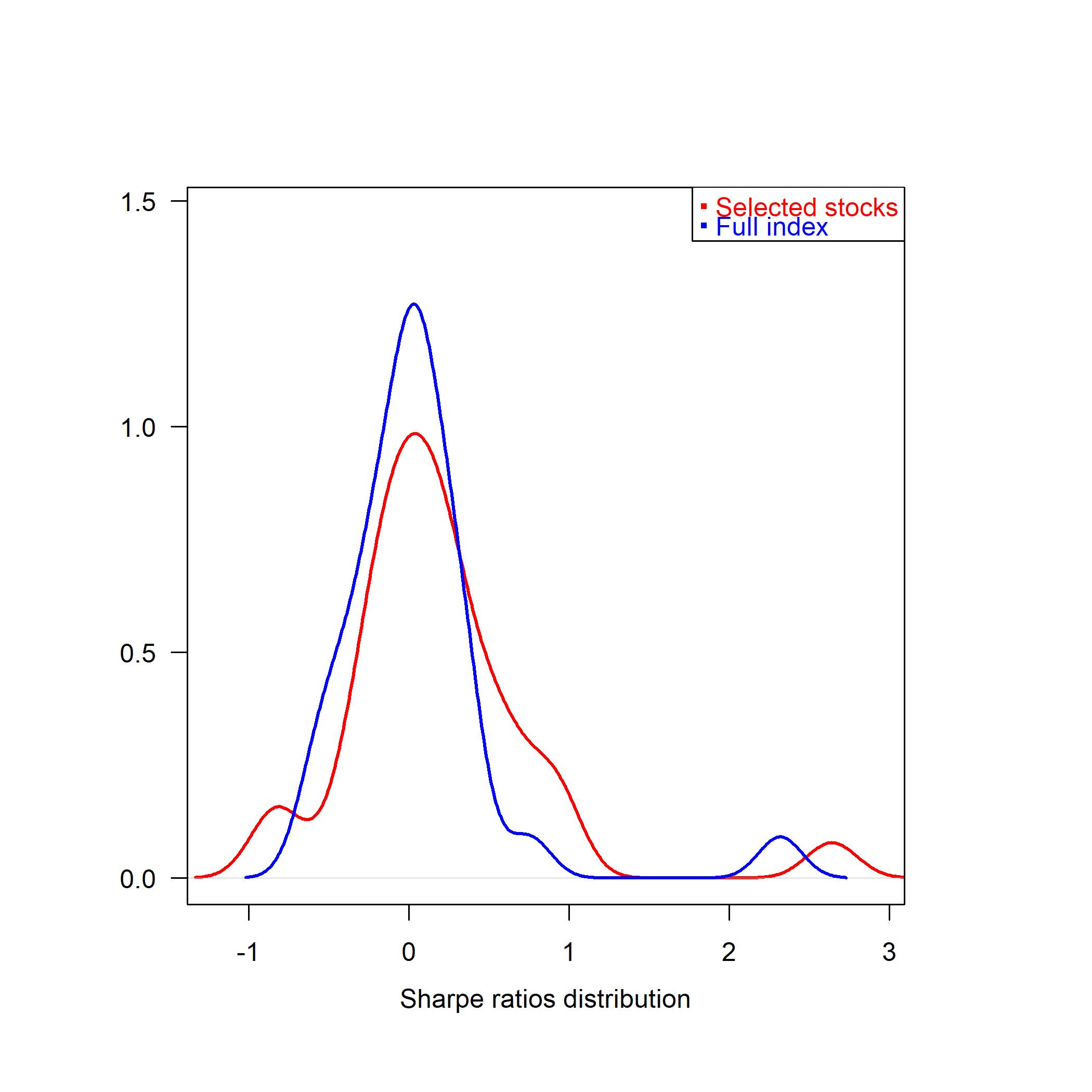}}
    \subfloat[]{\includegraphics[width=0.25\textwidth]{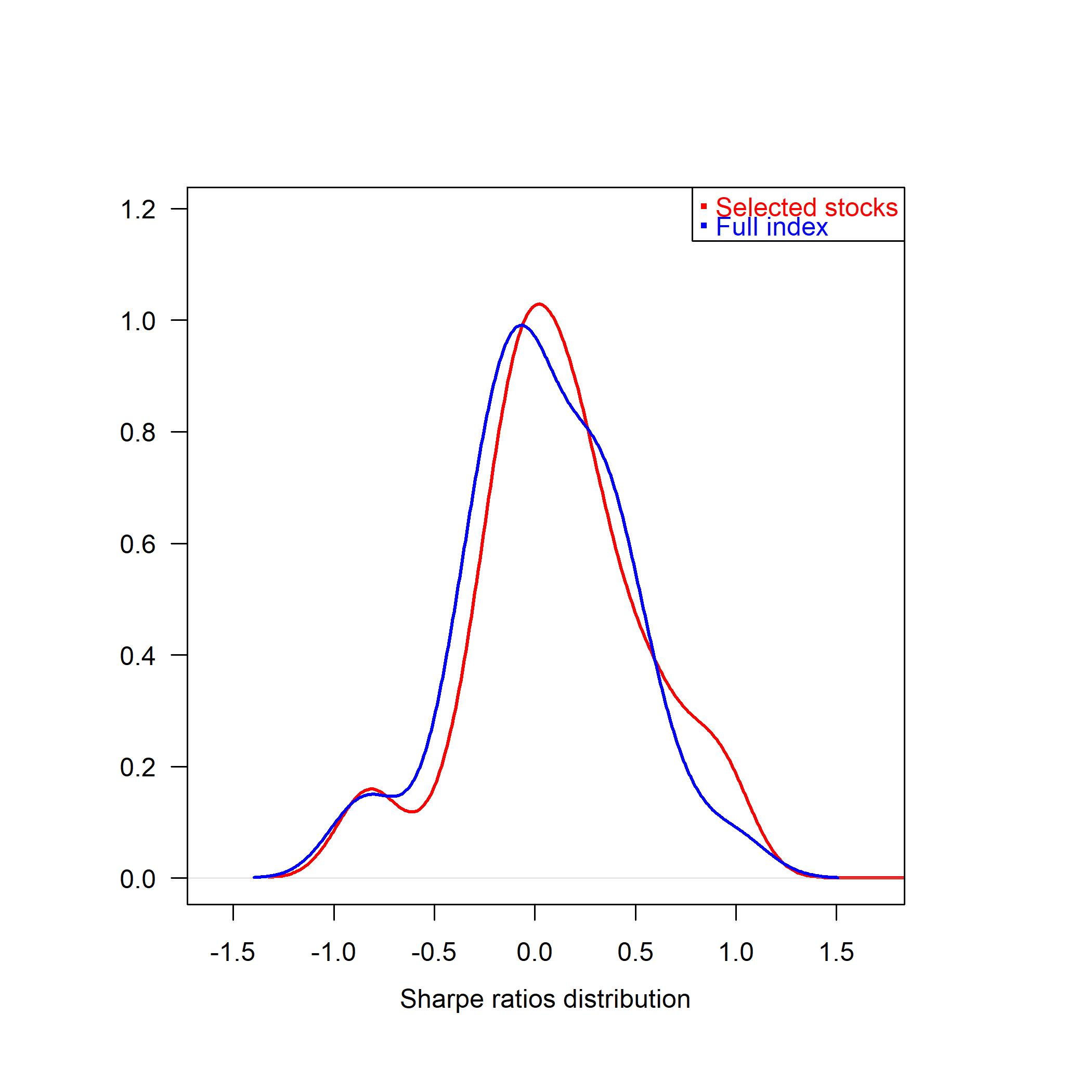}}\\
	\subfloat[]{\includegraphics[width=0.25\textwidth]{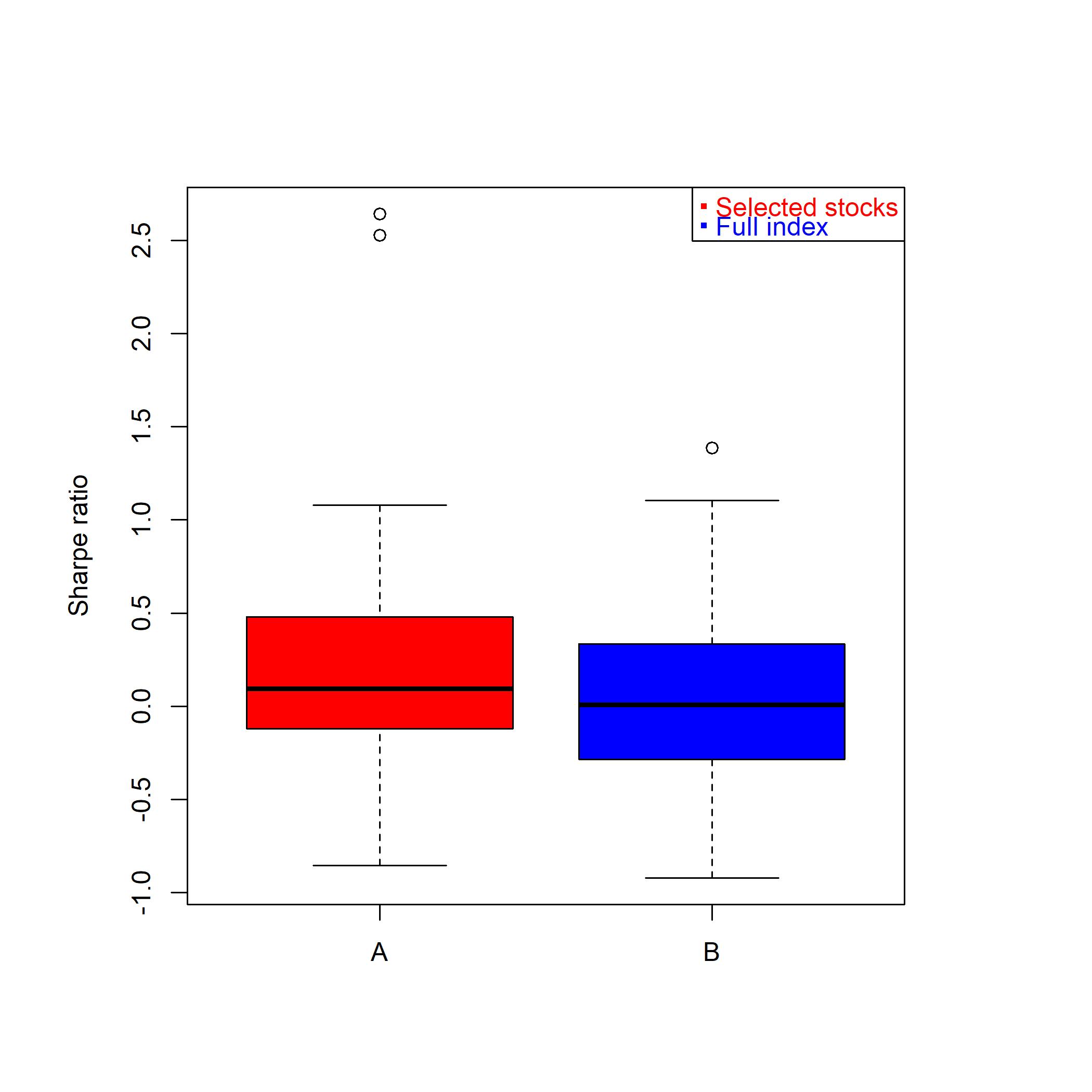}}
	\subfloat[]{\includegraphics[width=0.25\textwidth]{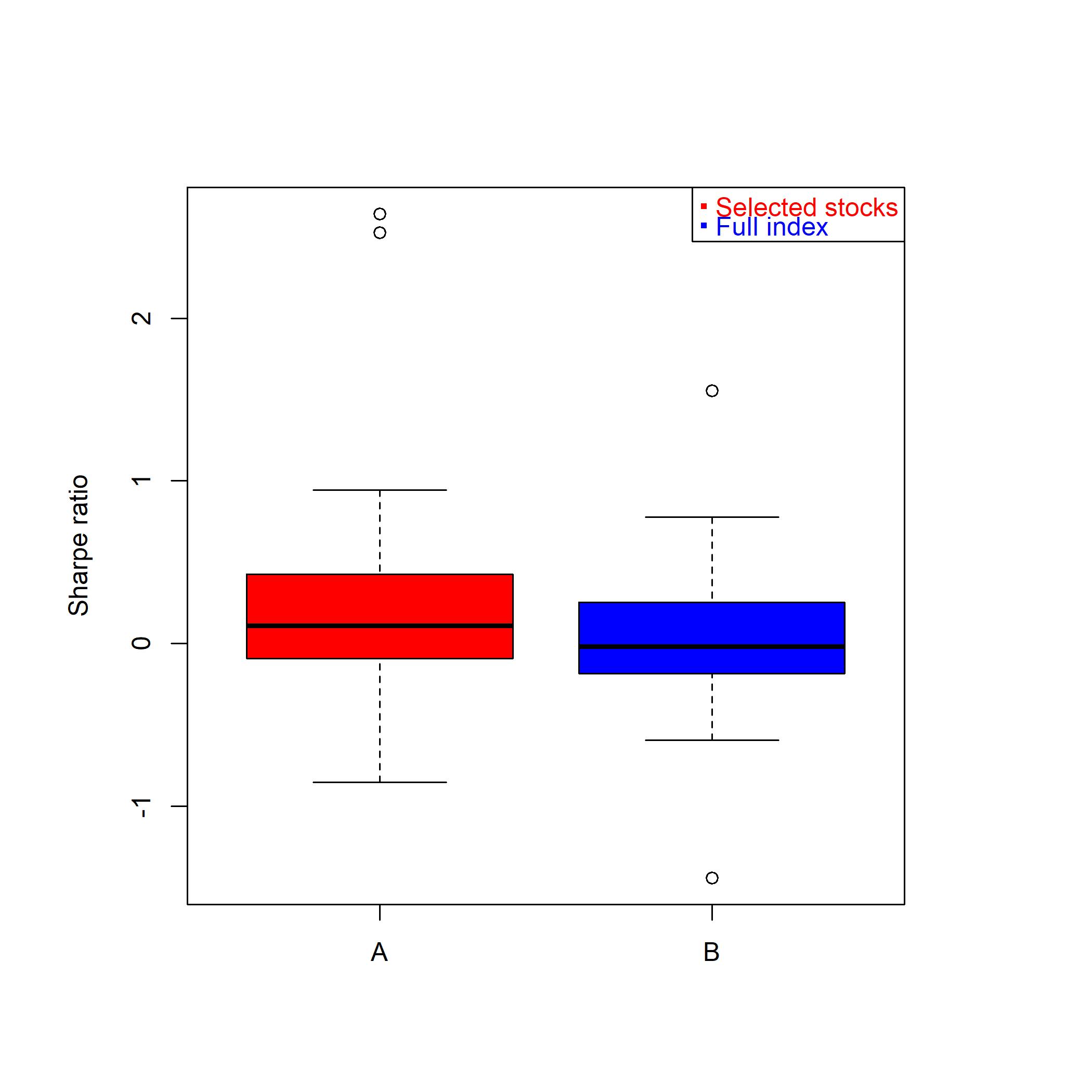}}	
	\subfloat[]{\includegraphics[width=0.25\textwidth]{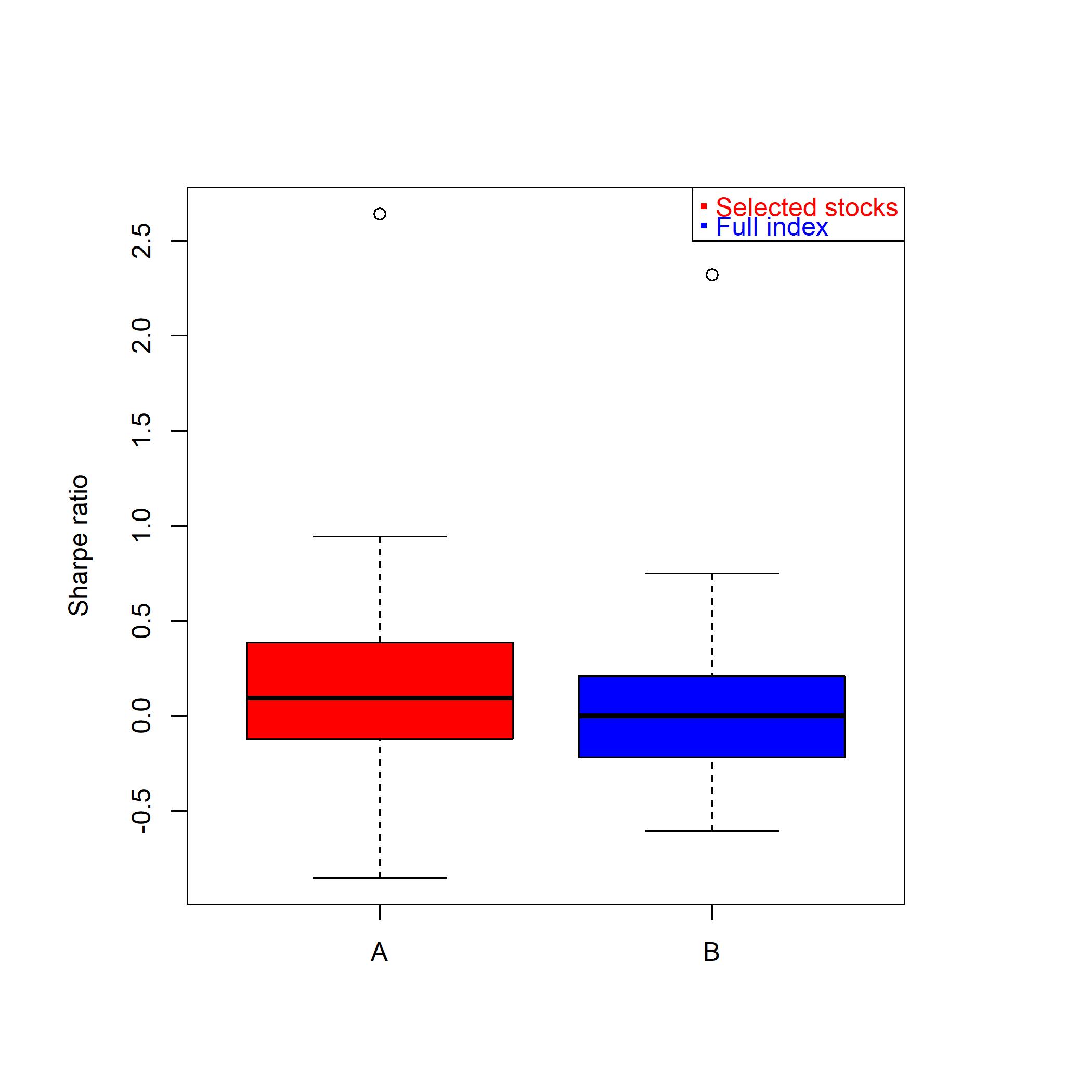}}
	\subfloat[]{\includegraphics[width=0.25\textwidth]{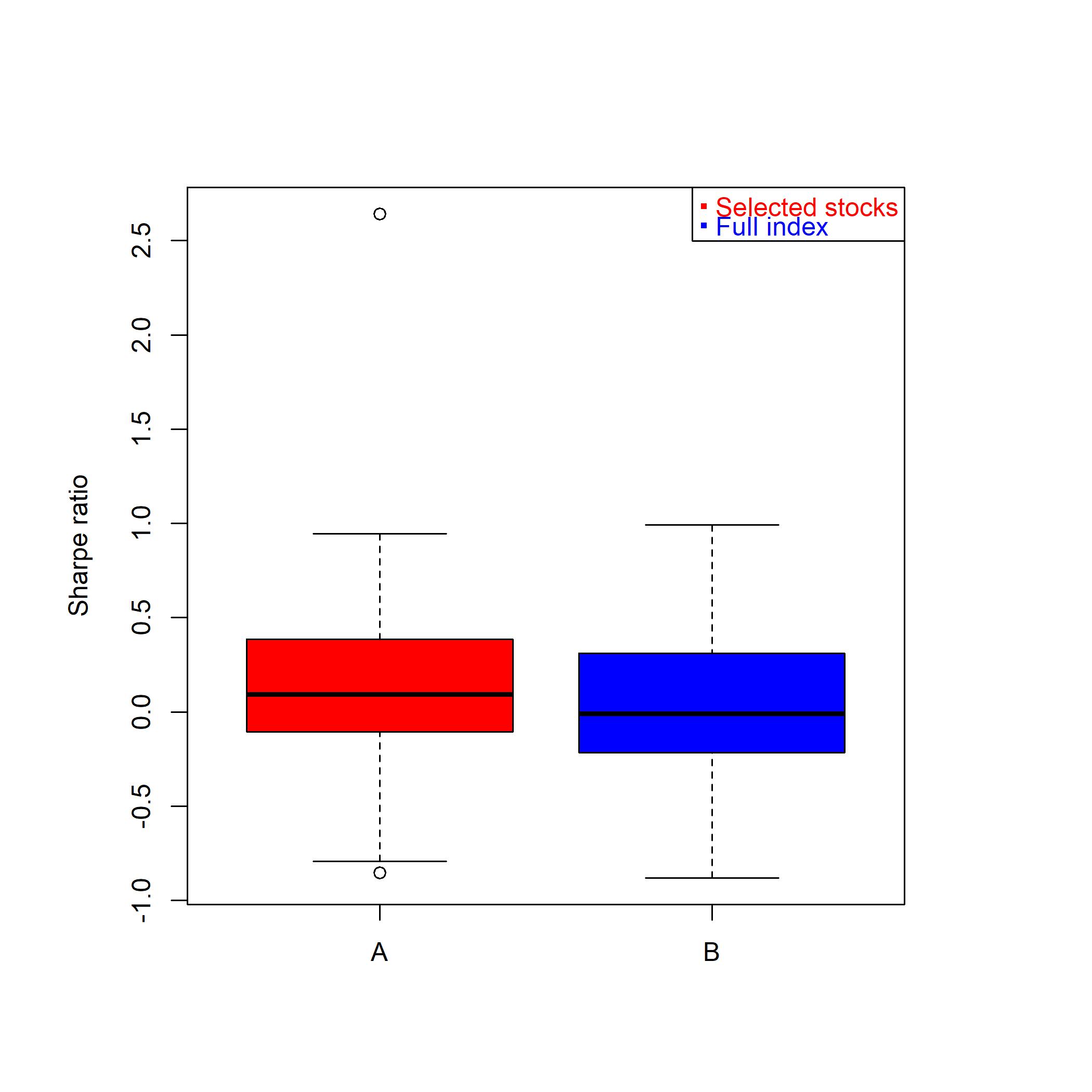}}
	\caption{Out-of-sample densities and box plots for the Sharpe ratios of two groups of selected stocks (in red) and the remainder of the dataset (in blue) for the NIKKEI index; $\tau_{G}=0.95$ for all plots.}
	\label{fig5}
\end{figure}

\begin{table}[h!]
\centering
\caption{Out-of-sample summary statistics for Sharpe ratios of selected stocks and global dataset across different threshold configurations for NIKKEI index.}
\label{tab:sharpe_ratios_nikkei_oos}
\begin{tabular}{ccccccc}
\toprule
\textbf{Panels} & \boldmath$\tau_L$ & \textbf{Group} & \textbf{Mean} & \textbf{Variance} & \textbf{Skewness} & \textbf{Kurtosis} \\
\midrule
(a), (e) & 0.45 & Selected & 0.2603452 & 0.5146761 & 1.716332 & 3.608807 \\
& & Global & 0.04530861 & 0.2191899 & 0.5939232 & 0.6505542 \\
\midrule
(b), (f) & 0.50 & Selected & 0.258769 & 0.5325863 & 1.751952 & 3.694765 \\
& & Global & 0.05657042 & 0.2661601 & 0.1199921 & 1.890272 \\
\midrule
(c), (g) & 0.55 & Selected & 0.187869 & 0.3785311 & 1.784601 & 5.63365 \\
& & Global & 0.05414324 & 0.2583662 & 2.644774 & 9.726742 \\
\midrule
(d), (h) & 0.60 & Selected & 0.1932738 & 0.3740916 & 1.802024 & 5.765645 \\
& & Global & 0.04606595 & 0.1607477 & -0.08739961 & 0.07603652 \\
\bottomrule
\end{tabular}
\label{table5}
\end{table}

Consider, as an example, the periods and stocks extracted from the NIKKEI index by the proposed out-of-sample approach with $\tau_{G} = 0.95$ and $\tau_{L} = 0.50$. In this case, the selected stocks exhibit a Sharpe ratio of $0.258769$, while the remaining stocks show a Sharpe ratio of $0.05657042$, indicating a positive gap in favor of the selected portfolio. We also observe that the skewness of the first distribution is higher than that of the second: $1.751952$ for the selected portfolio compared with $0.1199921$ for the remaining stocks. Even in the out-of-sample context, local balance proves to be an effective criterion for selecting securities with higher Sharpe ratios. This therefore empirically confirms the validity of the proposed methodology. 


\section{Conclusions}

The main idea proposed in this paper is to study deviations of local balance from global balance in a signed network to identify distinctive behaviors of specific nodes. This intuition is applied in a financial context by proposing an investment strategy that concentrates allocations in a few assets when two conditions are met: (i) a high level of global balance, indicating a financial crisis, and (ii) some nodes characterized by low levels of local balance. The rationale is that during periods of financial turbulence, standard diversification does not effectively reduce risk, and investors may instead consider concentrating their portfolios in a small number of assets that potentially outperform the market. During such crises, the risk associated with holding a concentrated portfolio is limited, as diversification opportunities reduce. The proposed criterion for selecting outperforming assets is based on the divergence between the global balance of the underlying signed network and the local balance of individual nodes. Experiments conducted on four datasets of real financial data confirm both the intuition and the effectiveness of the proposed investment strategy in outperforming the benchmark equally weighted portfolio, which approximates the market. The application demonstrates that meaningful results are obtained not only in a descriptive context but also in a predictive framework. Future research will focus on a more systematic investigation of these findings and on extending the proposed selection criterion to signed networks constructed from inputs other than the correlation matrix of returns.

\bibliographystyle{elsarticle-num}

\bibliography{References}

\end{document}